\documentclass[sigconf, screen]{acmart}

\AtBeginDocument{%
  \providecommand\BibTeX{{%
    \normalfont B\kern-0.5em{\scshape i\kern-0.25em b}\kern-0.8em\TeX}}}

\copyrightyear{2025}
\acmYear{2025}
\setcopyright{cc}
\setcctype{by}
\acmConference[CHI '25]{CHI Conference on Human Factors in Computing Systems}{April 26-May 1, 2025}{Yokohama, Japan}
\acmBooktitle{CHI Conference on Human Factors in Computing Systems (CHI '25), April 26-May 1, 2025, Yokohama, Japan}\acmDOI{10.1145/3706598.3713476}
\acmISBN{979-8-4007-1394-1/25/04}

\usepackage{subcaption} 
\usepackage{url}
\usepackage{verbatim}
\makeatletter
\g@addto@macro{\UrlBreaks}{\UrlOrds}
\makeatother
\usepackage{multirow} %tabular cells spanning multiple rows
\usepackage{color} %colour management
\usepackage{enumitem} %better  enumerate, itemize, and description

\usepackage{longtable}
\usepackage{xspace}

\newcommand{\m}{\textit{M=}}

\newcommand{\sd}{\textit{SD=}}

\newcommand{\F}[3]{$F({#1},{#2})={#3}$}

\newcommand{\p}{\textit{p=}}

\newcommand{\pminor}{\textit{p$<$}}

\newcommand{\chisq}{$\chi^2$}

\newcommand{\padjminor}{\textit{p$_{adj}<$}}
\newcommand{\padj}{\textit{p$_{adj}$=}}

\newcommand{\systemLong}{\textsc{Remote Operation Automated Driving Suite (ROADS)}\xspace}
\newcommand{\system}{\textsc{ROADS}\xspace}

\newcommand{\waypoint}{\textit{waypoint guidance}\xspace}
\newcommand{\trajectory}{\textit{trajectory guidance}\xspace}
\newcommand{\pathPlanning}{\textit{path planning}\xspace}

\newcommand{\interaction}{\textit{interaction concept}\xspace}
\newcommand{\interactions}{\textit{interaction concepts}\xspace}
\newcommand{\numberOfRequests}{\textit{number of requests}\xspace}

\newcommand{\AOASatisfying}{\textit{Satisfying}\xspace}
\newcommand{\AOAUsefulness}{\textit{Usefulness}\xspace}
\newcommand{\tlxScore}{\textit{TLX score}\xspace}
\newcommand{\SUSScore}{\textit{SUS score}\xspace}

\newcommand{\sumCurrentLaneDeviation}{\textit{absolute sum of lane deviation}\xspace}
\newcommand{\sumCurrentLaneDeviationAv}{\textit{average sum of lane deviation}\xspace}

\begin{document}

\title[ROADS: Systematic Comparison of Remote Control Interaction Concepts]{Introducing \textsc{ROADS}: A Systematic Comparison of Remote Control Interaction Concepts for Automated Vehicles at Road Works}

\author{Mark Colley}
\email{m.colley@ucl.ac.uk}
\orcid{0000-0001-5207-5029}
\affiliation{%
  \institution{Institute of Media Informatics, Ulm University}
  \city{Ulm}
  \country{Germany}
}
\affiliation{%
  \institution{UCL Interaction Centre}
  \city{London}
  \country{United Kingdom}
}

\author{Jonathan Westhauser}
\email{jonathan.westhauser@uni-ulm.de}
\orcid{0009-0003-2178-0448}
\affiliation{%
  \institution{Institute of Media Informatics, Ulm University}
  \city{Ulm}
  \country{Germany}
}

\author{Jonas Andersson}
\email{jonas.andersson@ri.se}
\orcid{0000-0002-6307-1960}
\affiliation{%
  \institution{RISE Research Institutes of Sweden}
  \city{Göteberg}
  \country{Sweden}
}

\author{Alexander G. Mirnig}
\email{alexander.mirnig@ait.ac.at}
\orcid{0000-0001-8848-0077}
\affiliation{%
  \institution{Department of Artificial Intelligence and Human Interfaces, University of Salzburg}
  \city{Salzburg}
  \country{Austria}
}
\affiliation{%
  \institution{Center for Technology Experience, AIT Austrian Institute of Technology}
  \city{Vienna}
  \country{Austria}
}

\author{Enrico Rukzio}
\email{enrico.rukzio@uni-ulm.de}
\orcid{0000-0002-4213-2226}
\affiliation{%
  \institution{Institute of Media Informatics, Ulm University}
  \city{Ulm}
  \country{Germany}
}

\renewcommand{\shortauthors}{Colley et al.}

\begin{abstract}
As vehicle automation technology continues to mature, there is a necessity for robust remote monitoring and intervention features. These are essential for intervening during vehicle malfunctions, challenging road conditions, or in areas that are difficult to navigate. This evolution in the role of the human operator—from a constant driver to an intermittent teleoperator—necessitates the development of suitable interaction interfaces. While some interfaces were suggested, a comparative study is missing. We designed, implemented, and evaluated three interaction concepts (\pathPlanning, \trajectory, and \waypoint) with up to four concurrent requests of automated vehicles in a within-subjects study with N=23 participants. 
The results showed a clear preference for the \pathPlanning concept. It also led to the highest usability but lower satisfaction. With \trajectory, the fewest requests were resolved. The study's findings contribute to the ongoing development of HMIs focused on the remote assistance of automated vehicles. 
\end{abstract}

\begin{CCSXML}
<ccs2012>
   <concept>
       <concept_id>10003120.10003121.10011748</concept_id>
       <concept_desc>Human-centered computing~Empirical studies in HCI</concept_desc>
       <concept_significance>500</concept_significance>
       </concept>
 </ccs2012>
\end{CCSXML}

\ccsdesc[500]{Human-centered computing~Empirical studies in HCI}

\keywords{remote operation, systematic comparison}

\maketitle

\section{Introduction}
%\todo{test as short paper? currently: 4950 words}
% Motivation
The road toward fully automated vehicles (AVs) is still paved with numerous challenges that must be overcome. Full automation, i.e., without the need for human intervention for regular intervention or fallback-readiness, is more readily achievable in controlled environments like airports or warehouses, where vehicle speeds and traffic stakeholders can be regulated. While AVs are already available in cities like Phoenix and San Francisco with state restrictions~\cite{wiredWaymoSF}, widespread adoption is not yet in sight. The technology is advancing rapidly and has anticipated significant advantages in safety, reliability, passenger comfort, and reducing economic and environmental costs~\cite{fagnant2015preparing}.

However, achieving highly automated driving (SAE level 4~\cite{SAE_J3016}) in mixed-traffic urban environments is challenging. When an environment can not be fully controlled, there will be situations that the automated system does not anticipate and will likely not possess the correct solution or exit strategy. This is why, during the transition phase towards full automation, for example, leveraged by delivery companies to benefit from anticipated lower costs, human safety drivers or operators are used to close this gap and be able to respond to unforeseen circumstances intuitively. Restrictions on the operational design domain (ODD), e.g., when local authorities prohibit the operation of AVs within road works~\cite{dean2022our}, further restrict the operation potential of fully automated AVs. To address these and similar challenges, remote operation systems have been proposed. These provide human monitoring and intervention capabilities, potentially in a one-to-many interaction setting, without requiring the physical presence of a human in the vehicle, where a remote operator (RO) from a control center guides the AV~\cite{kettwich_helping_2022, gafert_teleoperationstation_2022, gafert2023, kettwich_teleoperation_2021, majstorovic_survey_2022}. This was also shown in a pilot study referenced by \citet{vreeswijk_remote_2023}, which presents a real-world application. This strategy can apply to various vehicle types, including personal cars, vans, trucks, and buses.

% Problem
While several parts of remote operation, such as remote driving (\citet{tener_driving_2022}), HMI requirements (\citet{havoc}), and (technical) challenges and limitations (\citet{kang_augmenting_2018}), have been evaluated, a systematic comparison of user interfaces (UIs) for RO intervention when a vehicle needs assistance has not yet been conducted. In addition, it is commonly assumed that a teleoperator will monitor -- and potentially have to intervene on -- more than a single vehicle \cite{gafert2023}. Yet, the question of how many assistance events one operator can handle, given a specific interaction design, is not well studied for regular assistance scenarios~\cite{kneissl2020}.
% Position your approach
Therefore, we implemented an integrated remote operation system called \systemLong. \system allows the RO to use the three concepts of \waypoint, \trajectory, and \pathPlanning, as described in the overview of \citet{majstorovic_survey_2022}.
In the remote operation scenario, between one and four AVs have to be controlled and steered through road works, as this could constitute the end of an ODD. In a within-subject design with N=23 participants, we found that the participants preferred \pathPlanning. The worst rating was given to \trajectory.

\textit{Contribution statement: } This work contributes (1) the implementation of three interaction concepts for remote operation of AVs, which will be openly available to the broader community, and (2) results of a within-subjects study with N=23 participants evaluating the interaction concepts with one to four parallel requests showing that multiple requests are feasible with degraded performance and that \pathPlanning was preferred.

\section{Related Work}

This work builds on prior work on teleoperation and, in general, on cooperation between humans and automation.

\subsection{Scenarios for Teleoperation}
\citet{kettwich_helping_2022} analyzed possible scenarios for teleoperation in road traffic. These scenarios were gathered via interviews and observations with control center staff, video analyses from real-world road events, and interviews with AV safety operators. \citet{kettwich_helping_2022} categorized use cases into eight clusters, including passenger interaction, technical malfunctions, and interactions between the environment, the RO, the AV, and others. 

In our study, the scenario consists of road works too complex or out of the legal boundary for our AV to handle independently and, therefore, is out of its ODD, relating to the unmapped construction site~\cite{kettwich_helping_2022}. %They further propose that a vehicle could navigate through the road works on its own using road-site units as support. However, in our scenario, this option is not feasible as it surpasses the vehicle's ODD.

\subsection{Teleoperation Concepts}
\label{ssec:conceptsofteleop}
AV teleoperation intensities can be divided into three categories. \citet{bogdoll_taxonomy_2022} defined %collected multiple terms with the same usage to define a universal taxonomy for remote operation systems. The terms 
\textit{remote monitoring}, \textit{remote assistance}, and \textit{remote driving} in ascending order of the interaction intensity (see also \citet{andersson2024navigating}). \textit{Remote monitoring} refers to a one-sided stream of information from the vehicle to the RO. \textit{Remote assistance} requires the RO to aid the AV in its decisions but not take over to perform actual driving tasks, contrary to \textit{remote driving}. \citet{bogdoll_taxonomy_2022} define \textit{remote driving} "as 'real-time performance of part or all of the DDT [dynamic driving task] and/or DDT fallback [...] by a remote driver.'"~\cite[p. 2]{bogdoll_taxonomy_2022}, while directly controlling the vehicle "in the form of steering and throttle/brake commands"~\cite[p. 10]{bogdoll_taxonomy_2022}.

Using \citet{bogdoll_taxonomy_2022}'s taxonomy, \citet{majstorovic_survey_2022} surveyed prior work. They split \textit{remote driving} and \textit{remote assistance} into concepts. We evaluated concepts of \textit{trajectory guidance} (building a path by dragging the mouse cursor), \textit{waypoint guidance} (generate a connected path by placing single points), and \textit{interactive path planning} (user chooses one of multiple system-offered paths) and compared them. We exclude other concepts as they are either strongly tied to \textit{remote driving} or too use case specific. We specifically exclude the concepts of explicit \textit{remote driving} due to their limitations in teleoperation (see Section~\ref{sec:limitations}). %, as well as to provide a fair comparability between the analyzed concepts.

Although \citet{majstorovic_survey_2022} classify \textit{trajectory guidance} as a form of \textit{remote driving}, based on the practical implementation we are proposing, the fluid transition between those categories and the corresponding definition in the taxonomy by \citet{bogdoll_taxonomy_2022}, for further discussion, we categorized it as \textit{remote assistance}.
Additionally, the interaction type for actions in remote assistance is very different compared to active control due to the different levels of interaction abstraction (direct movement vs., e.g., point guidance) and the event-driven interaction requests in remote assistance~\cite{majstorovic_survey_2022}. 
%Furthermore, there are specific use cases in the teleoperation of vehicles~\cite{kettwich_helping_2022} where remote assistance is unavailable due to, for example, the unavailability of parts or subsystems crucial for autonomous driving. \citet{bogdoll_taxonomy_2022} also restrict the usage domain of \textit{remote assistance} to SAE-Level 4 vehicles. 
%Thus, when to use which modality is still to be examined and discussed in future work. These considerations should also consider the applied limitations to both categories (e.g., latency requirements and part or subsystem availability).

\citet{brecht2024evaluation} evaluated several teleoperation concepts (direct control, shared control, trajectory guidance, waypoint guidance, collaborative planning, and perception modification) from an ego perspective with eight experts in the automotive and teleoperation industry. Such experts might not be the final users of these teleoperation concepts, so we had participants without a background in these industries. They found that ``a holistic teleoperation system should be composed of implementations of the Shared Control, Collaborative Planning and Perception Modification concepts''~\cite[p. 639]{brecht2024evaluation}.

%Complementary to this study, 
\citet{colley_systematic_2022} conducted a literature survey regarding the "final 100 meters problem"~\cite{colley_systematic_2022}, which relates to the variability of the final destination in a journey based on user preferences and technical limitations. The survey was conducted on "possible interaction modalities and modes"~\cite[p. 2]{colley_systematic_2022}, creating an overview of possibilities to communicate with the vehicle. %The findings include different experimental methods for controlling the vehicle directly. Derived from this, a study was conducted with VR simulation, which, among other objectives, compared different methods of interaction modalities for actively controlling an AV. 
The results of a VR study indicated a preference for a steering wheel, followed by using a joystick. %As \citet{colley_systematic_2022} point out, the preference towards the steering wheel could be due to the familiarity with this modality from the driving experience. % contrary to the general novelty towards the other modalities in this aspect.
%Our work follows a similar approach but in a different context. 
While \citet{colley_systematic_2022} evaluated interaction concepts from inside the AV, we compare concepts for remote assistance. This opens new possibilities and interaction perspectives (e.g., handling multiple requests simultaneously or simplified representations of the environment).

\subsection{Human Factors}
\label{sec:human_factors}
\subsubsection{Human Factors in Remote Operation}
\citet{kari_human_2021} identified the human factors related to remote operations. Although they focus on remote \textit{ship} operations, many factors apply in an automotive context. Especially situational awareness (SA)~\cite{graf_improving_2020, gafert_teleoperationstation_2022, kettwich_teleoperation_2021, majstorovic_survey_2022}, the lack of physical sensing and soundscape~\cite{tener_driving_2022} and the implied delays~\cite{cummings_concepts_2020,tener_driving_2022,kang_augmenting_2018} are important for ROs. %This is especially true for driving tasks but also for remote assistance, which requires consideration in the design of remote control interfaces for AVs.

%Another aspect to be considered is presented by \citet{hashimoto_human_2022}. 
Additionally, \citet{hashimoto_human_2022} raised the question of what the maximum feasible count N of vehicles managed concurrently is. Further challenges involved are determining the difficulty levels when one RO has to grasp the situation for multiple vehicles in case of emergencies or determining "how many additional operators (up to N-1 carts) are required when one cart is in the takeover?"~\cite[p. 6]{hashimoto_human_2022} (cart $\widehat{=}$ vehicle).
Additionally, switching between vehicles and switching costs shall be considered~\cite{hashimoto_human_2022}. Thus, our study addresses the applicability of one operator managing multiple vehicles concurrently. It takes performance measurements on parallel incoming vehicle requests, tied with the currently used interaction modality. To support parallel request handling, our proposed UI provides an additional secondary display for a selected vehicle to display its front camera to monitor the vehicle.

\subsubsection{Organization and Roles in Remote Operation Centers (ROCs)}
\citet{schrank_roles_2021} proposed a "first conceptual step" for an organizational structure for ROCs. While being a preliminary concept, it lays a foundation for future discussions. This proposal defines three central roles, the Remote Coordinator (RC), the Remote Driving Operator (RDO), and the Remote System Operator (RSO), along with multiple Peripheral Roles that support and complement core operations. 
The primary task of an RC is to maintain an overview and monitor all incidents and how they are handled. RDOs are responsible for remotely controlling individual AVs, while RSOs manage a fleet of AVs.

\subsection{Interface Requirements}
\label{ssec:interfacereq}
Several studies were conducted to determine the requirements for remote control interfaces for AVs, for example, the HAVOC Project~\cite{havoc}\footnote{\url{http://tinyurl.com/havocproject}; Accessed by: 29.11.2024}. 15 participants monitored 10 automated trucks and had to assist in five occurring events actively. Two of those events were active control tasks conducted with a steering wheel, pedals, and a computer screen. %Several remarks (user needs) were collected concerning remote monitoring, assistance, and remote driving. 
The feedback focused on the lack of information to consciously control the vehicle, including the need for a 360° overview of the surroundings and the uncertainty of when the AV can take over again, which we aimed to resolve in our prototype. To achieve this within the interface, we included a projected birds-eye view of the vehicle, added information on collision avoidance detection, and notified the RO when the vehicle took over again.
%These aspects, therefore, were considered when designing the apparatus for the user study, as partly described in section~\ref{sec:study}.

Comparably, \citet{tener_driving_2022} studied remote driving.  %Although remote driving has different interface requirements than remote assistance, some overlap exists.
The requirements we identified and adapted for our use case concerning remote assistance included emphasizing the takeover reason, giving contextual road information, and giving AI suggestions for operations~\cite{tener_driving_2022}. We included automated suggestions only in \pathPlanning for differentiation. %However, we kept automated suggestions reserved for \pathPlanning as this already embodies a selection from given suggestions and otherwise would break the fair comparison of the pure interaction concepts.

\citet{gafert_teleoperationstation_2022} used a demonstrational VR remote driving setup, whose interface was based on the requirements of \citet{graf_user_2020}.
To gather these requirements, \citet{graf_user_2020} conducted 18 interviews. Half of them had some remote controlling experiences. %, although not necessarily with vehicles. 
80 requirements were collected. In a second interview with participants from industry, these requirements were clustered and rated, resulting in a final set of 20 requirements (e.g., 360-degree remote view, vehicle position, vehicle issues, traffic rules), where we adopted the relevant ones. % While these requirements were collected for a remote driving background, not all may apply to remote assistance, especially regarding the necessity rate. Nevertheless, we identified and adapted the relevant requirements for our interface, such as a 360-degree (top-down) view from the vehicle, information concerning the reason for the takeover, projected control actions, and vehicle speed.

%Similar to our study and apparatus, 
\citet{kettwich_teleoperation_2021} designed a Human Machine Interface (HMI) for the teleoperation of vehicles in a public transport control center context. %They gathered data from a study with 12 participants, all employees in public transport control centers in Germany. 
They wanted to receive "a first impression on whether the development is in the right direction, particularly whether its overall setup is valid"~\cite[p. 19]{kettwich_teleoperation_2021}.
The findings included that the proposed HMI establishes a "suitable interface design for the teleoperation of highly automated vehicles in public transport"~\cite[p. 15]{kettwich_teleoperation_2021}. However, some suggested improvements were related to overwhelming information. Participants dealt with 6 monitors and operated with a keyboard and mouse and an additional touchscreen. Thus, we created a more minimalist design to maintain focus on the most important elements for the given task (see Section~\ref{sec:study}).

Further, \citet{kettwich_teleoperation_2021} defined seven HMI evaluation criteria: (1) Features, (2) Information, (3) SA, (4) Usability, (5) User Acceptance, (6) Attention, (7) Capacity. We used questionnaires to evaluate our prototypes. We used the SUS~\cite{brooke1996sus} to measure user usability, the van der Laan acceptance scale~\cite{VANDERLAAN19971} to measure User Acceptance, and NASA-TLX~\cite{hart1988development} to measure user capacity. Additionally, participants were asked for liked and missed features. However, we did not include dedicated scales for measuring the remaining criteria: Information, SA, and Attention, but instead relied on logged data to provide recognition to them. %, sufficient to address our research question.

\subsection{Limitations of Remote Controlling}
\label{sec:limitations}
While controlling AVs remotely brings many advantages, limitations also exist. \citet{cummings_concepts_2020} and \citet{tener_driving_2022} raise critical issues addressing the remote driving of AVs. Delays concerning the technical side of communication and the human side of reorientation and reaction time narrow the applicability of remote operations. Therefore, \citet{cummings_concepts_2020} point out that requiring an RO to take over the vehicle at high speeds in a short time frame in the range of seconds would be unsafe. Taking these limitations into account, the expected safe situations that remain open for a takeover are those in which the vehicle is traveling at slow speeds, less than 10 mph~\cite{cummings_concepts_2020}, or has already pulled to the shoulder by performing a DDT as specified by the SAE~\cite{SAE_J3016} in a critical situation.
The technical delay in communication was further elaborated by \citet{kang_augmenting_2018}. They present a case study further investigating the imposed delay by comparing the effect of different levels of resolutions in LTE~\cite{seo_lte_2016} and Wi-Fi~\cite{song_wi-fi_2017}, which represent "wireless networks used by vehicles"~\cite[p. 22]{kang_augmenting_2018}.
% \url{https://www.telekom.com/de/konzern/details/5g-geschwindigkeit-ist-datenkommunikation-in-echtzeit-544496#:~:text=Die%20durchschnittliche%20Latenz%20im%20LTE,Zeit%20von%20drei%20Millisekunden%20erzielt.}
Their results show that higher resolutions, i.e., higher image sizes, lead to higher image latencies. %Subsequently, two optimization techniques are proposed. The first is to dynamically adjust the intervals of the intra-coded frames (I-frames) relative to the predicted frames (P-frames) based on the different vehicle dynamics. The second is to minimize the transmitted data by transmitting only the differences between two 3D maps on both sides.%Ergebnis

Concerning the given limitations on remote driving operations, we propose that the effects of limitations on remote assistance are less intensive and, therefore, bearable. As the task of speed and brake control is delegated to the vehicle, many involved security issues caused by latency and human factors (see  Section~\ref{sec:human_factors}) are minimized or entirely mitigated. Additionally, enabling the RO to plan vehicle routes for limited distances within remote assistance also enables them to assist multiple vehicles simultaneously, contrary to basic remote driving.
Therefore, our study focuses on remote assistance to take advantage of the benefits an SAE 4 vehicle presents to mitigate security risks.

%Statement re. research/dev. gap that we address

\section{User Study}\label{sec:study}

We systematically explored three interaction concepts based on a previous overview by \citet{majstorovic_survey_2022}: \waypoint, \trajectory, and \pathPlanning. Additionally, we were interested in the effect of multiple parallel requests. While the apparatus allows up to seven parallel requests, internal pilot tests before the main study showed that four parallel requests are already highly demanding. Therefore, the independent variables were the \interaction and the \numberOfRequests. 
We did not include overview maps and technical details of the requesting vehicle. While these are relevant data for a real-life application, for our research question (RQ), these were irrelevant due to the focus on a self-contained scenario without references to other locations.
The two RQs that guided this study were:

\begin{quote}
    RQ1: What are the effects of different \textit{interaction concepts} on potential remote operators in terms of (1) usability, (2) acceptance, (3) performance, (4) efficiency and effectiveness? \newline
    RQ2: What are the effects of scaling the \numberOfRequests with the \interaction?
\end{quote}

\subsection{Apparatus}\label{sec:study-apparatus}

\begin{figure}[ht]
    \centering
    \includegraphics[width=.5\textwidth]{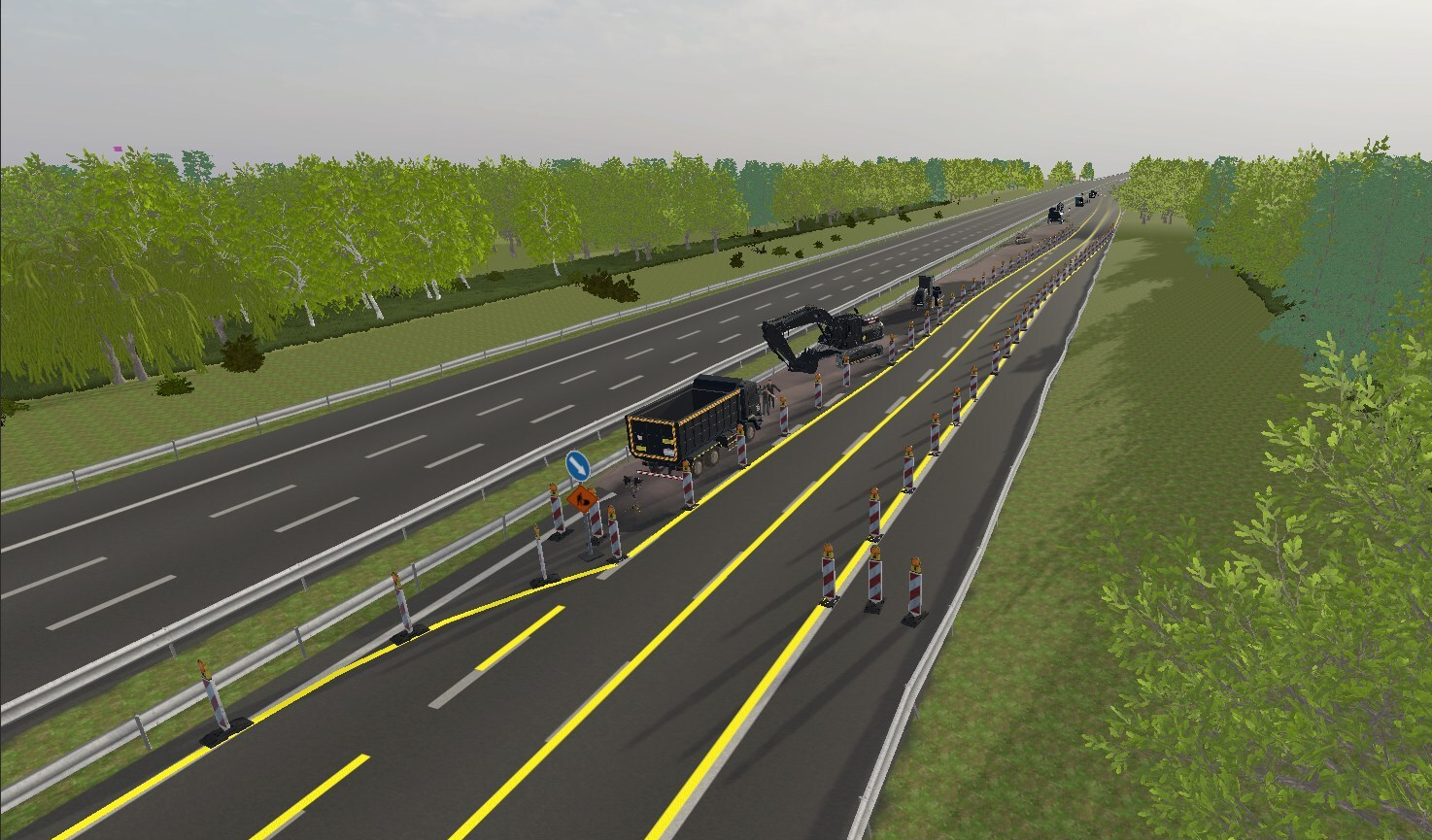}
    \caption{Scene overview of the road works.}
    \label{fig:overview-scene}
    \Description{A screenshot from inside the simulation demonstrating what the approached road works looks like.}
\end{figure}

The apparatus comprises a Unity application for the driving and interaction environment (see \autoref{fig:overview-scene}). We designed a standalone application using \href{https://unity.com/}{Unity} 2022.3.7. 
This application simulates one or more AVs approaching road works. These road works are also mirrored to avoid participants becoming too accustomed to the environment without changing the actual task's difficulty. In the scenario, there are three lanes with other vehicles. The driving environment aligns with previous research on takeovers on highways~\cite{colley2021takeover}. We used the Mobile Traffic System asset\footnote{\url{https://assetstore.unity.com/packages/tools/behavior-ai/mobile-traffic-system-194888}; Accessed on 14.08.2024} to simulate traffic behavior. The traffic and AV speed at the start of the interaction is $\approx$120km/h. 

The traffic alongside the assisted AV behaved according to German traffic regulations, which prohibit overtaking within this road works area due to the present lane markings. Consequently, if a request is neglected (e.g., waiting for instructions) for too long, it will cause a traffic jam in its lane.

When a path is processed, the AV begins to follow it without prior effectiveness checks (e.g., we did not check whether this is the fastest path or whether the path would lead to a successful exit of the construction site). Therefore, planning a suitable path is entirely delegated to the RO, while the AV retains responsibility for proper path following, including collision avoidance and cruise control. This follows the application requirements of at least an SAE-Level 4 vehicle to perform \textit{remote assistance}, as specified by \citet{bogdoll_taxonomy_2022}.

To resolve an incoming request, a total distance of 600 meters had to be traversed, of which the vehicle had an initial path of 200 meters already assigned when issuing the request, which could optionally be overwritten by the RO. This path distance is implied by the specified view range of our remote assistance view, which is set to 200 meters, starting from the vehicle. This aligns with the minimum range requirement for \textit{LiDAR} sensors, as specified by \citet{lidarRange}. Current \textit{LiDAR} systems also already meet this requirement as \citet{stateOfTheArt2021} mention. 
To safely travel the distance required to resolve the request without changing the initial zoom level, an RO would have to make a minimum of ten separate path inputs for \trajectory, seven for \waypoint, or three for \pathPlanning. However, zooming was enabled, therefore, these values are just for reference.

The RO was given 120 seconds to resolve all incoming requests in the study. When multiple requests were to be assisted, all of them came in at the same time.

\begin{figure}
    \centering
    \includegraphics[width=.35\textwidth]{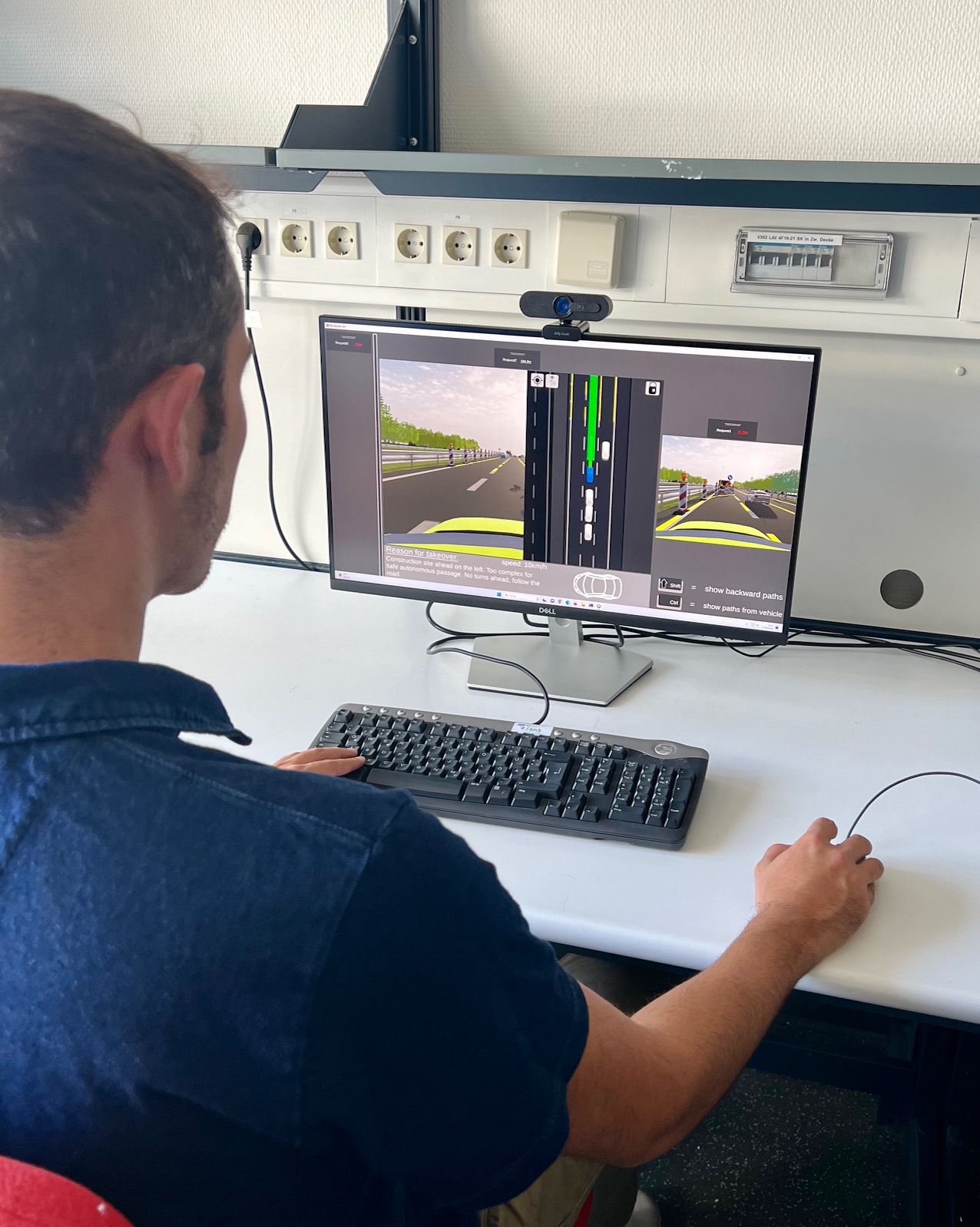}
    \caption{Workspace of the participants. The image shows the monitor used for the experiment with a mounted webcam.}
    \label{fig:participant}
    \Description{A camera image showing a person from behind in front of a computer monitor, currently handling vehicle requests in the apparatus, using a mouse and keyboard.}
\end{figure}

Regarding hardware, we used a QHD 27-inch monitor (i.e., the Dell S2721D) and a standard cable mouse and keyboard (see \autoref{fig:participant}). In this work, we deliberately omitted technical limitations of RO such as latency to test the \interaction effectively.

\subsection{Procedure}
Each participant experienced \textbf{12} conditions, representing a 3 $\times$ 4 design (\textit{interaction concept} with three levels: \trajectory, \waypoint, and \pathPlanning and \numberOfRequests with values ranging from one to four). Each condition (i.e., combination of \interaction and \numberOfRequests) had one trial. Therefore, every participant had to handle 30 requests in total.  

Each session started with a brief introduction and signing of the consent form. Afterward, participants had time to get accustomed to the systems. Each participant was introduced to all system capabilities, including the user interface and the interaction concepts. Furthermore, each participant was encouraged to familiarize themselves with the system and particularly with the interaction concepts until becoming confident with them. For this purpose, the same scenario was used as in the real conditions with two concurrent requests. The conditions were then presented in counterbalanced order using a Latin square. The conditions with the same \interaction were not placed together. No participant mentioned any frustration by constantly switching interaction modalities.
The introduction to the capabilities of the system was given as follows: 

\begin{quote}
    \textit{You are a remote operator for automated vehicles that are not capable to operate in all possible conditions. Therefore, if they encounter scenarios in which they cannot or are not allowed to operate, you are asked to take control from afar. Depending on the condition, you will use different interaction concepts to guide the automated vehicle through the scenario.}
\end{quote}

After all conditions, a demographic questionnaire was filled out by the participants. On average, a session lasted 80 min. Participants were compensated with 15€.

\subsection{User Interface}

\subsubsection{Request Management}

\begin{figure*}[ht]
    \centering
    \includegraphics[width=0.75\textwidth]{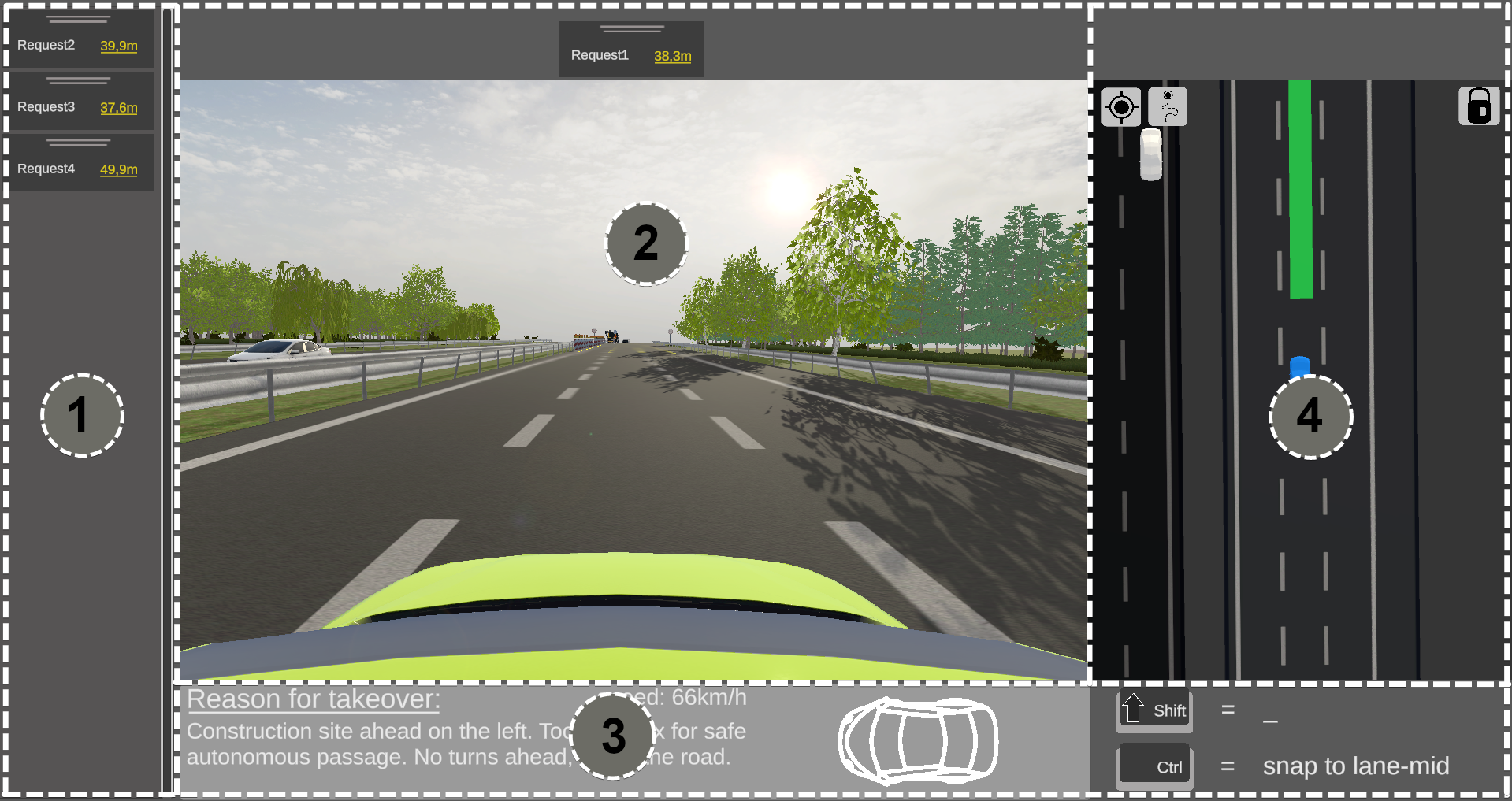}
    \caption{The main screen of the application. Here, four requests have to be handled in parallel. UI elements overview: (1) request list, (2) main request camera, (3) information panel, (4) remote assistance input panel.}
    \label{fig:application}
    \Description{A screenshot of the user interface. One request is active as the main request. On the left side, there is the request list. The center displays the front camera of the main request, with its control input window on the right side of the screen. The user interface elements are assigned numbers that are referenced.}
\end{figure*}

At the beginning of the interaction, the left panel shows the requests of the vehicles. The middle is still empty and at the bottom, there are fields for information on the emerging requests (see \autoref{fig:application}). The possible keystrokes on the right apply to the interaction concept used.

The requests can be dragged to the center or to the right edge via drag-and-drop. The areas for this are highlighted.
Up to two requests can be active at the same time; while only one of them can be controlled (the large one in the middle), the other one (right at the edge) can only be observed.

In the lower panel, the reason for the required takeover is described. Just above it is the speed of the vehicle, and to the right of it, a representation of a vehicle, from which the front and rear are illuminated in yellow as soon as something is near the vehicle. This indicates that the vehicle cannot drive in that direction.
%The information about this represents the vehicle in each case, which is shown large in the center.

In the input panel for the interaction, there are three buttons. 
From left to right, first, the "Vehicle Focus" button. When pressed, the view constantly follows the vehicle. 
The second is the "Path End Focus". Here, the view jumps to the end of the path as long as this button is active. The focus is lost as soon as the operator moves the view by pressing and dragging the left mouse button.
Third is the "lock button". When this is active, the operator restricts the manual dragging direction of the view to only forward or backward movement. Otherwise, all directions are possible.
One can also drag opened requests back to the list on the left anytime.
The goal is to answer the requests in such a way that the vehicle can drive independently again. The vehicle will announce itself by means of a text in the middle as soon as this is the case. The request disappears shortly thereafter.

\subsubsection{Interaction Concepts}

In general, starting and stopping were handled by the AV as long as collision avoidance did not trigger and was no task of the RO. Compared to previous work, e.g., by \citet{brecht2024evaluation}, our interface provided a bird-view visualization compared to an ego perspective. Importantly, these concepts also work \textbf{without} clear lane separation; therefore, simplifications like buttons for ``move along'' or ``switch lane'' maneuvers were avoided.

\begin{figure*}[ht!]
\centering
\small
    \begin{subfigure}[c]{0.485\linewidth}
        \includegraphics[width=\linewidth]{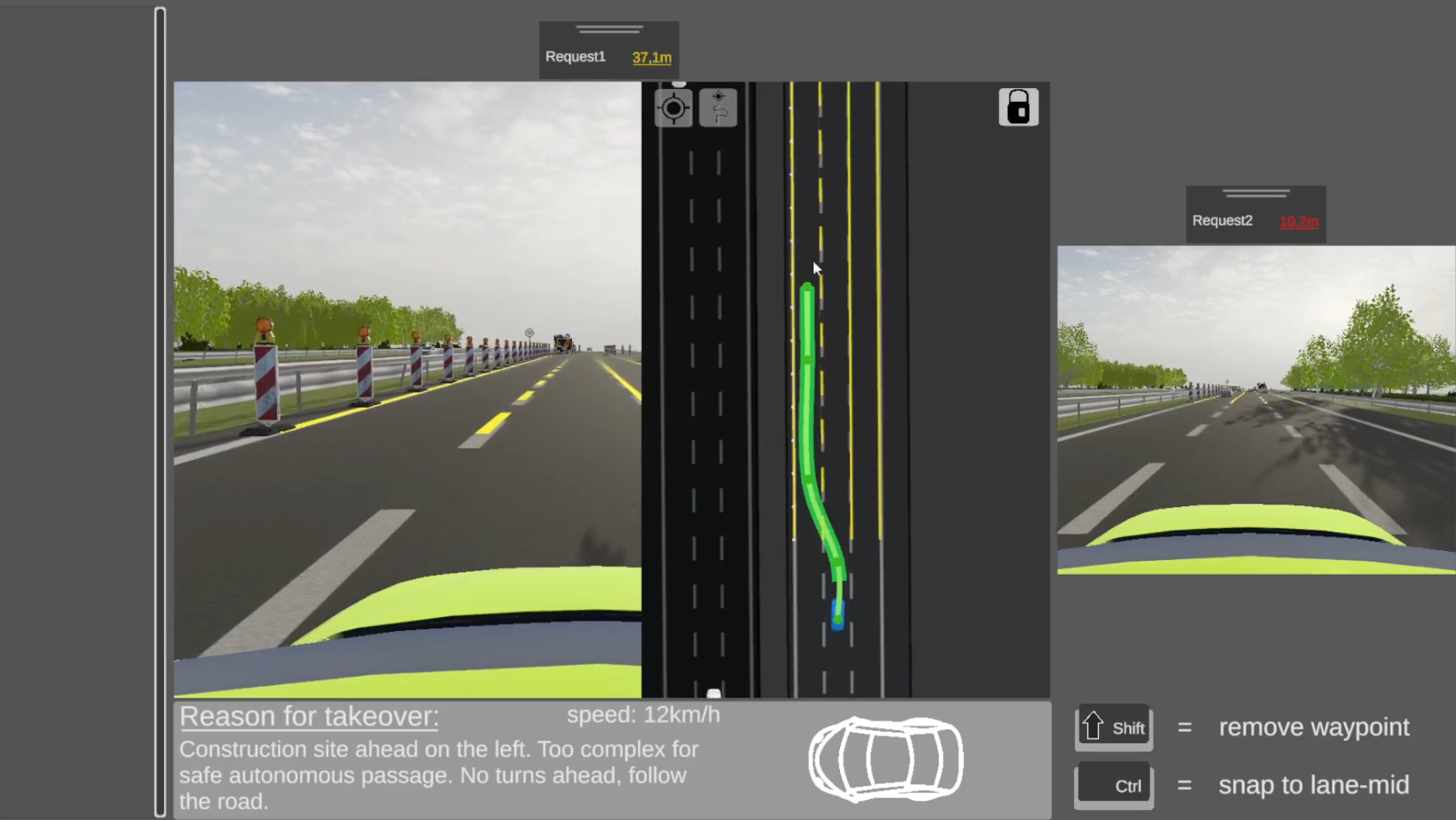}
        \caption{Waypoints. 4 individual waypoints (shown in dark green) are automatically connected. A second request is shown to the right.}\label{fig:waypoint}
        \Description{A screenshot of the UI with the waypoint interaction concept active. The user has currently placed 4 waypoints which generate a connection line, for the vehicle to follow. Here, for displaying purposes, a second request is also opened to monitor it.}
    \end{subfigure} 
        \begin{subfigure}[c]{0.485\linewidth}
        \includegraphics[width=\linewidth]{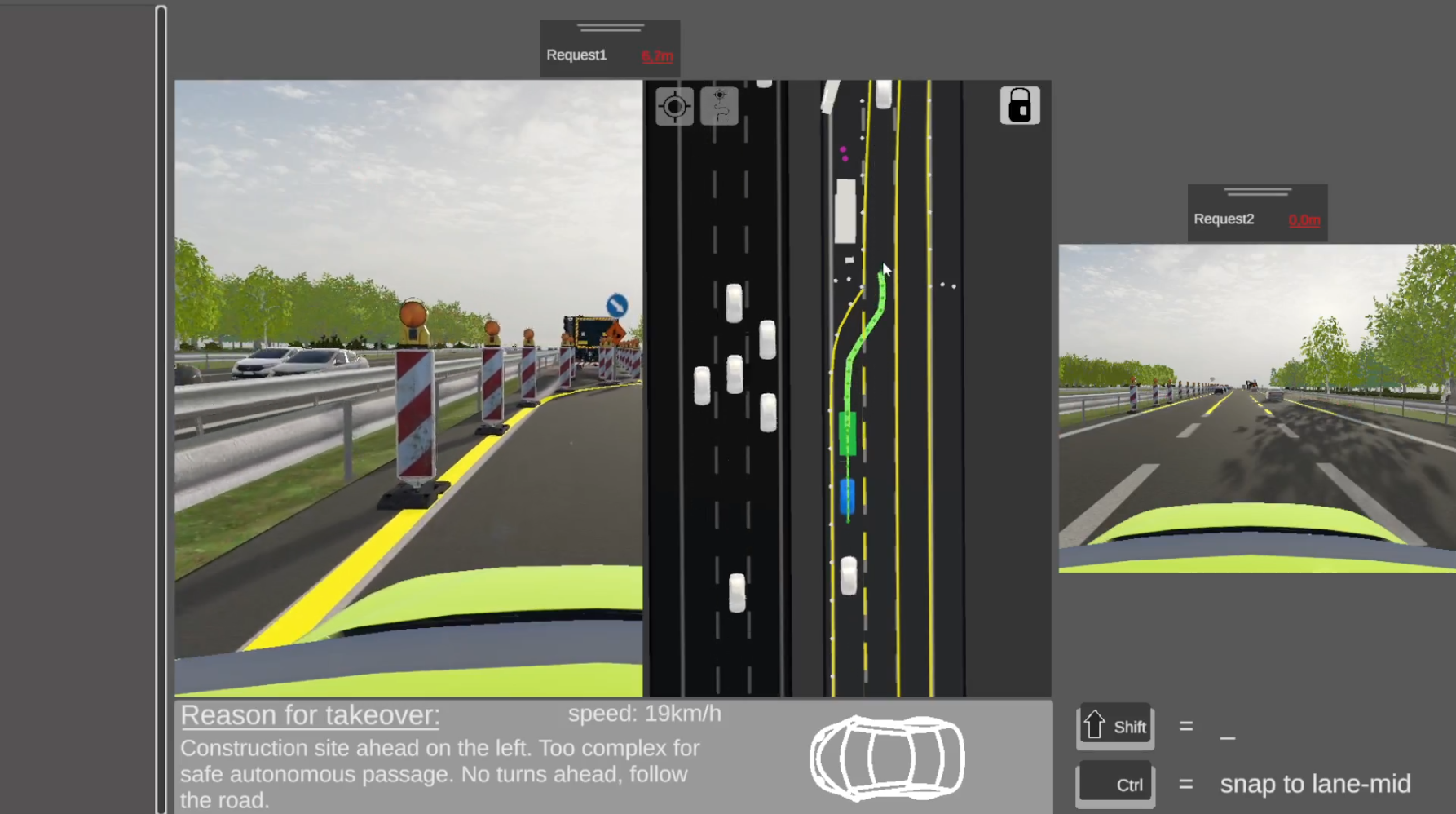}
        \caption{Trajectories. By moving the pressed mouse cursor, points (shown in dark green) are combined into a trajectory.}\label{fig:trajectory}
        \Description{A screenshot of the UI with the trajectory interaction concept active. The user has currently drawn a line, implying a lane change with it. Here also, a secondary request is opened}
    \end{subfigure} 
        \begin{subfigure}[c]{0.485\linewidth}
        \includegraphics[width=\linewidth]{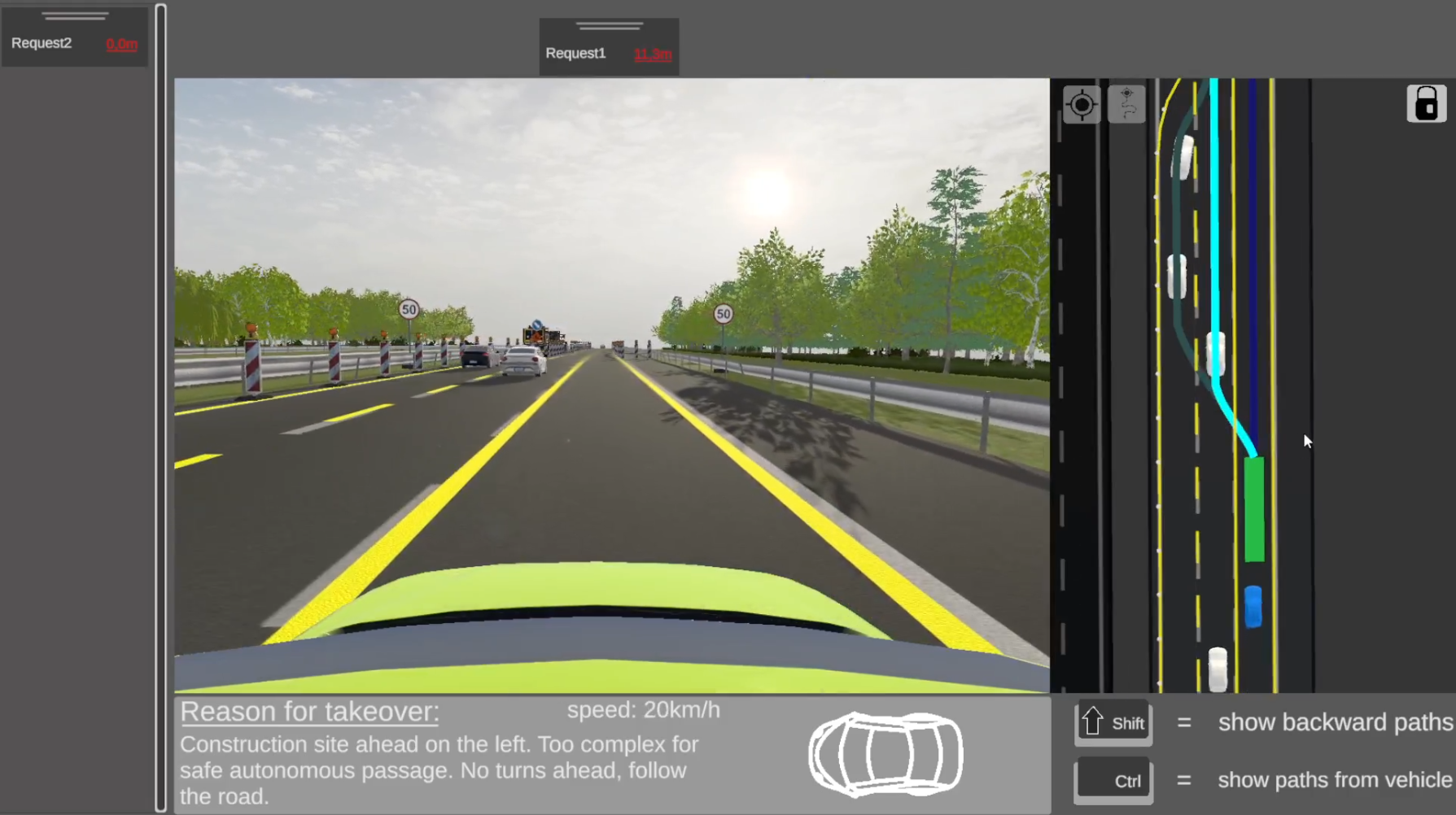}
        \caption{Path planning. Three potential paths are visible and color-coded. A second request is in the queue don't he left.}\label{fig:pathplanning}
        \Description{A screenshot of the UI with the path planning interaction concept active. The user has the option to choose between three distinct paths shown as colored lines in front of the vehicle}
    \end{subfigure}
    \caption{Overview of employed interaction concepts for remote operation. The ego-vehicle is shown in blue. The green part shows the future trajectory that is already planned.}
    \label{fig:overview-ui}
   \Description{A group of three screenshots displaying the three different interactions and what usage of them looks like visually.}
\end{figure*}

\paragraph{Waypoint Guidance}

The user can form the path by sequentially clicking the right mouse button and the resulting points (see \autoref{fig:waypoint}; also called waypoint guidance by \citet{brecht2024evaluation}). Points that would form an internal angle of $<=90$° will result in no waypoint set, as this maneuver would not be possible by a vehicle. 90° turns are possible with advanced steering systems\footnote{\url{https://www.youtube.com/watch?v=uZas2YCV3JY}; Accessed: 29.11.2024}. Importantly, the final movement of the AV was always determined by the Polarith AI asset\footnote{\url{https://assetstore.unity.com/packages/tools/behavior-ai/polarith-ai-pro-movement-with-3d-sensors-71465}; Accessed on 08.12.2024}. This incorporates calculations for semi-realistic movements. 
The RO can add waypoints between others and move points after pressing the mouse. In addition, pressing the Shift key and clicking the mouse can delete a waypoint again. Holding the control key when placing and moving points sets the position to the center of a lane, similar to the \trajectory interaction.

\paragraph{Trajectory Guidance}
The RO can draw the path by dragging the mouse (see \autoref{fig:trajectory}; also called trajectory guidance by \citet{brecht2024evaluation}). The right mouse button must be pressed during this process. This generates equidistant waypoints. Hectic movements in opposite directions are blocked. These are detected by taking the next three consecutive waypoints and forming an (imaginary) triangle. If the internal angle at the second point is $>=90$°, the input is blocked. 
When the mouse button is released, the newly drawn path is accepted and the vehicle starts to follow it. This interaction also enables to extend and fully or partially replace paths.

What happens in the end when drawing new paths depends on the distance of the start and end points to the last drawn path:
\begin{itemize}
\item \textit{Extension}: If only one point of the new path is close (i.e., 3.5m) to the existing path, the old path is extended. The extension direction depends on the distances to the start and end points and the vehicle involved.
\item \textit{Replacement}: If both the start and end points of the new path are close to the old one, the segment in between on the old path is replaced with the new path.
\item \textit{Parallel Replacement}: If the new path is close enough to the old one and runs parallel, the parallel part is replaced with the new path. This rule is rarely used and has a low threshold to prevent interference with extensions. 
\end{itemize}

Holding down the control key automatically draws in the middle of the lane next to the cursor, simplifying the process of drawing along the roadway when no special maneuvers, such as changing lanes or driving outside regular lane boundaries, are required.

\begin{figure}[ht!]
    \centering
    \includegraphics[width=.225\textwidth]{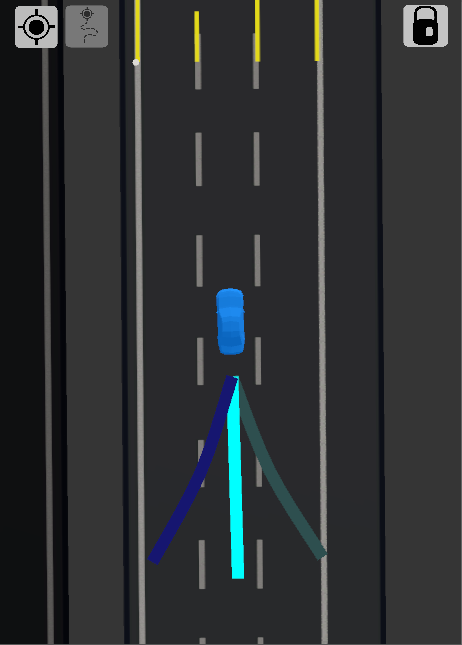}
    \caption{Backward paths in \pathPlanning. Three options are possible, as for forward paths. For visual distinction, other colors are used.}
    \label{fig:backwardPaths}
    \Description{A screenshot that shows how the reverse paths are generated for the remote operator.}
\end{figure}

\paragraph{Path Planning}
The user can set the path by selecting from suggested paths with the right mouse button (see \autoref{fig:pathplanning};  called collaborative planning by \citet{brecht2024evaluation}). In the evaluated scenario, up to three paths (two while driving through the construction site) are continuously generated depending on the situation (i.e., the number of lanes). By holding the shift key, one can select from three static generated reverse paths, as seen in \autoref{fig:backwardPaths}. These do not align with any roadway shapes and represent a fallback option to reposition in case of blocking objects in front of the vehicle or required orientation adjustments. This interaction concept includes the assumption that the AV will still be able to plan possible paths but either might not be allowed to execute them due to a legal boundary or the confidence in the applicability might be too low. Waymo claims to ``identify [...] stop signs greater than 500 meters away''~\cite{waymo2020fifth}, thus, we cautiously showed trajectories up to 185m in front of the AV.

%\subsubsection{Failure Handling}

\subsection{Participants}
We computed the desired sample size for the main experiment via an a-priori power analysis using G*Power \citep{faul2009statistical}.
To achieve a power of .8, with an alpha level of .05, theoretically, 21 participants should result in an anticipated medium effect size (\textit{Effect Size f}=0.27) in a repeated measures ANOVA. 

We recruited N=23 participants (Mean age = 23.9, SD = 2.60, range: [18, 28]; Gender: 34.8\% women, 60.9\% men, 4.35\% non-binary; Education: High school, 56.52\%; College, 43.48\%). Participants played video games on average \m{12.91} (\sd{15.43}) hours per week. On 5-point Likert scales (\textit{1 = Strongly Disagree} --- \textit{5 = Strongly Agree}), participants showed high interest in AVs (\m{4.57}, \sd{.51}), believed AVs to ease their lives (\m{4.22}, \sd{.85}), but were skeptical about whether they become reality by 2033 (\m{3.57}, \sd{1.34}).
Participants had no prior expertise in remote control.

\subsection{Measurements}\label{sec:measures}
\textit{Objective dependent variables:} During each session, the system logged the eye gaze with 60 Hz for the four areas of interest: request panel, info panel, main panel, and secondary panel. We also logged the current lane deviation in meters and mouse movements (only after the third participant due to technical problems) in cm with 10 Hz.

\textit{Subjective dependent variables:} After each condition, we measured the Task Load Index via the NASA-TLX~\cite{hart1988development} on 20-point scales and usability via the system usability scale (SUS)~\cite{brooke1996sus} which is a 10 item questionnaire with five response options: from Strongly agree to Strongly disagree. These items are: ``I think that I would like to use this system frequently.,
``I found the system unnecessarily complex.'' ,
``I thought the system was easy to use.'', 
``I think that I would need the support of a technical person to be able to use this system.'',
``I found the various functions in this system were well integrated.'',
``I thought there was too much inconsistency in this system.'',
``I would imagine that most people would learn to use this system very quickly.'',
``I found the system very cumbersome to use.'',
``I felt very confident using the system.'',
``I needed to learn a lot of things before I could get going with this system.''.
Finally, we asked participants about their acceptance using the van der Laan acceptance scale~\cite{VANDERLAAN19971} consisting of nine semantic differentials rating a system. It measures acceptance on the two subscales Usefulness  (averaging the five items: Useful $\leftrightarrow$ Useless, Bad	$\leftrightarrow$	Good, Effective	$\leftrightarrow$	Superfluous, Assisting	$\leftrightarrow$	Worthless, Raising Alertness	$\leftrightarrow$	Sleep-inducing) and Satisfying  (averaging the items: Pleasant	$\leftrightarrow$	Unpleasant, Nice	$\leftrightarrow$	Annoying, Irritating	$\leftrightarrow$	Likeable, Undesirable	$\leftrightarrow$	Desirable).

After all conditions, participants rated their preferences regarding the interaction concepts from greatest (\textit{ranking = 1}) to lowest (\textit{ranking = 3}). Open questions regarding improvement proposals were also asked (i.e., ``Why did you rate them that way?'' and ``Please describe further positive aspects and ideas for improvement.'').

\section{Results}

%1. eye gaze ANOVA?
%2. Mouse movements reporting?
%Sankey Diagram?

\subsection{Data Analysis}
Before every statistical test, we checked the required assumptions (e.g., normality distribution).
For non-parametric data, we used the ARTool package by \citet{wobbrock2011art} as the typical ANOVA is inappropriate with non-normally distributed data and Holm correction for post-hoc tests (using Dunn's test). The procedure is abbreviated, as in the original publication, with ART.
R in version 4.4.1 and RStudio in version 2024.09.0 was employed. All packages were up to date in December 2024.

\subsection{Eye Gaze}

\begin{figure*}[ht!]
\centering
\small
    \begin{subfigure}[c]{0.49\linewidth}
        \includegraphics[width=\linewidth]{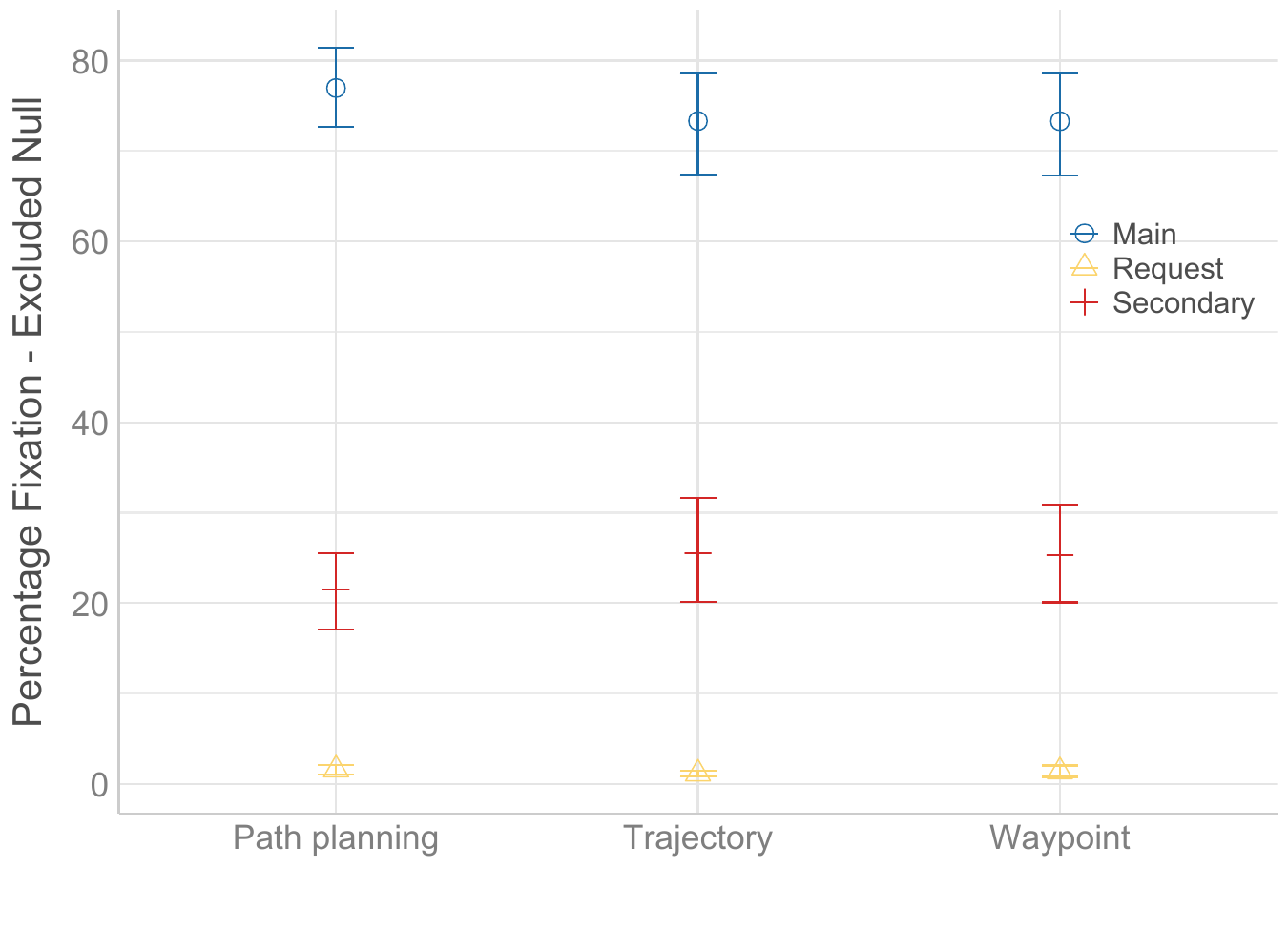}
        \caption{}\label{fig:fixations_interaction}
        \Description{A graph displaying the eye gaze area percentages, split between the different interaction concepts.}
    \end{subfigure}
    \begin{subfigure}[c]{0.49\linewidth}
        \includegraphics[width=\linewidth]{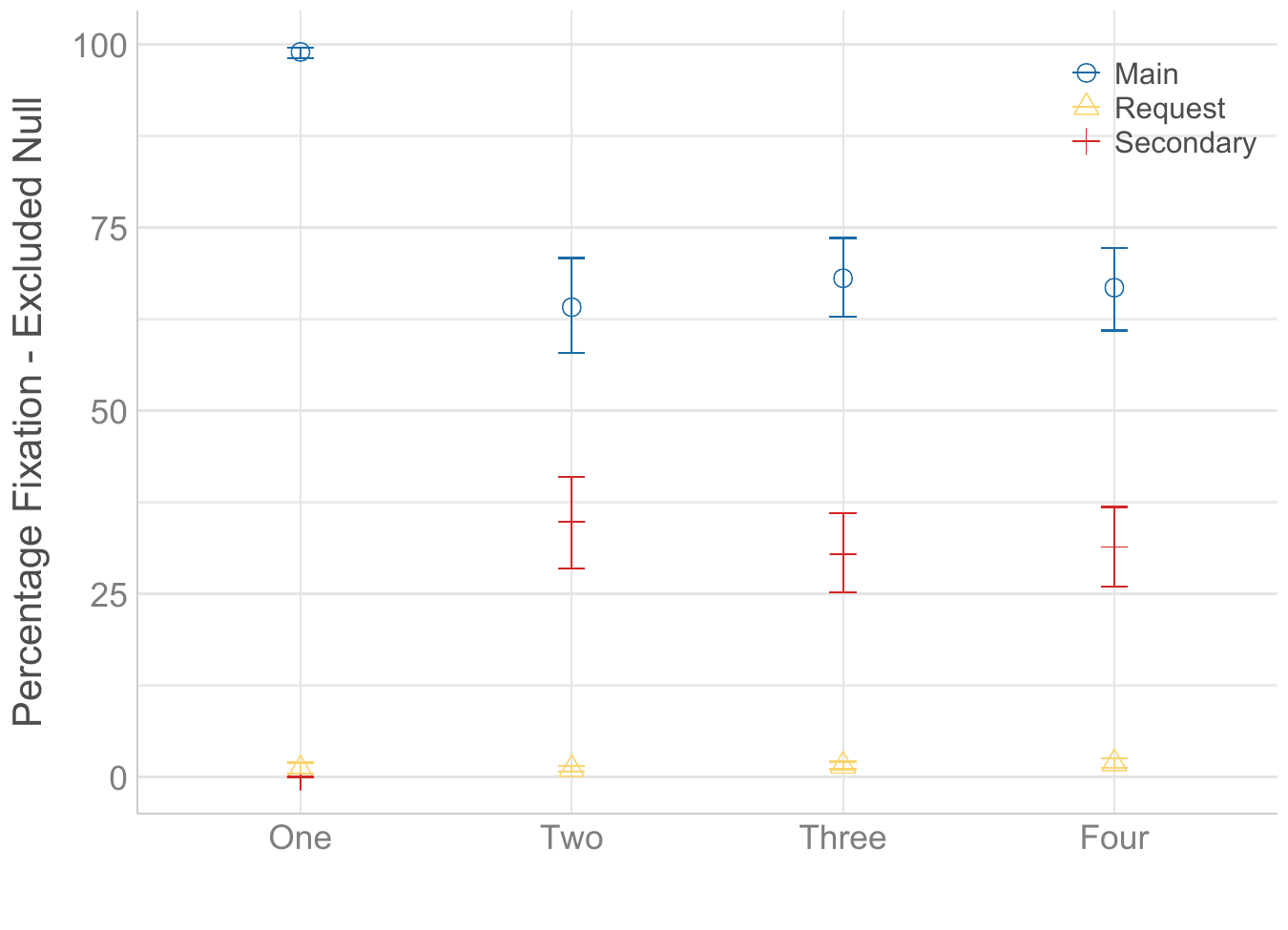}
        \caption{}\label{fig:fixations_number}
        \Description{A graph displaying the eye gaze area percentages, split between the number of total requests in this scenario}
    \end{subfigure} 
    \caption{Fixations regarding the \interaction (a) and the \numberOfRequests (b). Displayed are the gaze areas in percentages.}
    \label{fig:fixations}
   \Description{A group of two graphs displaying information on the eye gaze areas in percentages. The areas are split between the main panel for the main request, the secondary panel for a secondary request, and the request list. Subfigure A shows the different interaction concepts. Subfigure b shows the number of requests. }
\end{figure*}

The descriptive eye gaze data in \autoref{fig:fixations} shows clearly that the main panel was always viewed the most. The actual request panel received very little attention from the participants. If multiple requests were present, participants did focus $>25\%$ of their visual attention on the secondary panel (see \autoref{fig:fixations_number}).

\subsection{Number of Missed Requests}

\begin{figure}
    \centering
    \includegraphics[width=.45\textwidth]{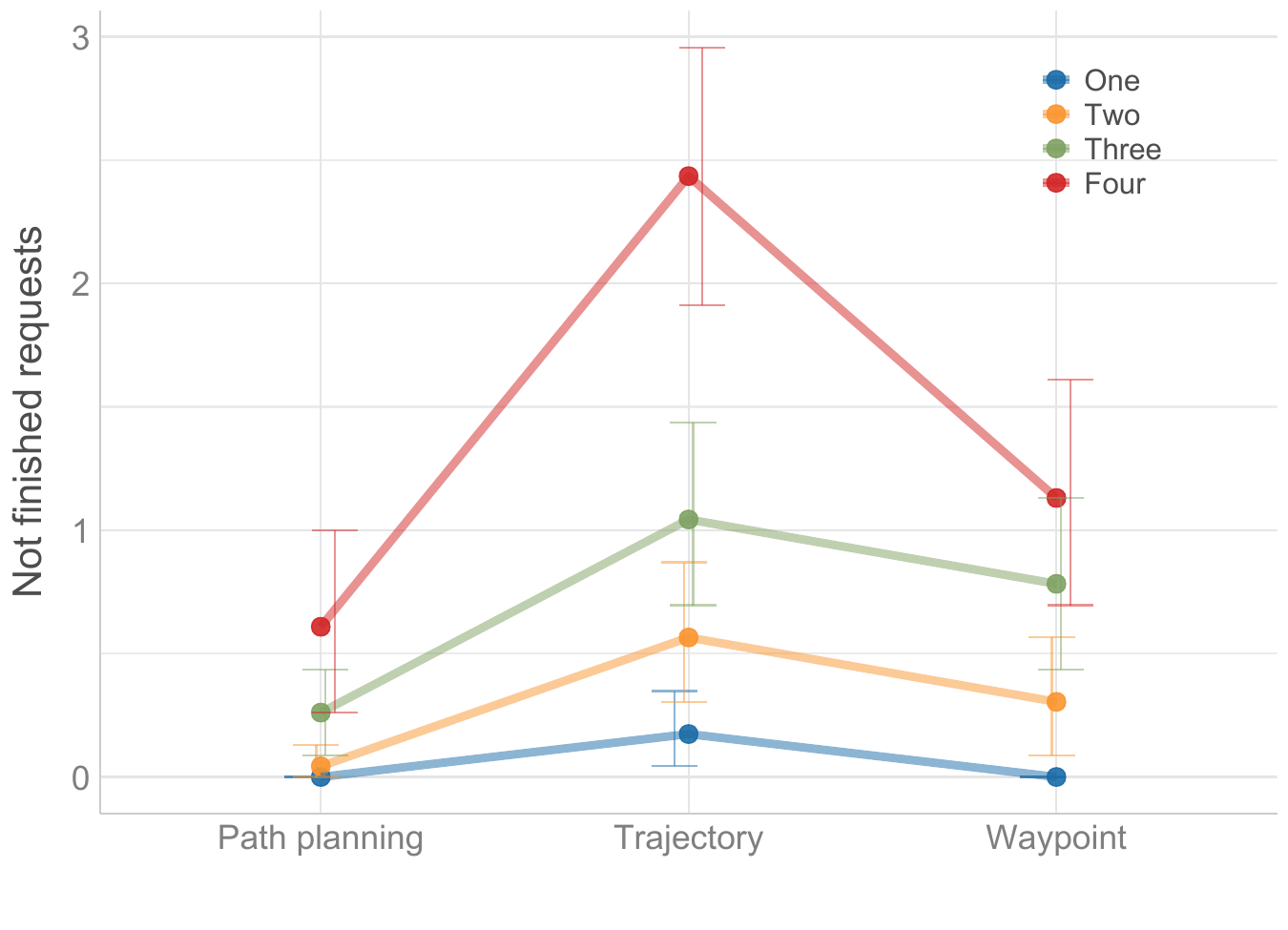}
    \caption{Number of missing requests per \interaction.}
    \label{fig:missingRequests}
    \Description{A graph of the number of not successfully handled requests, split between the three interaction concepts. For each concept, this is evaluated separately for 1 to 4 total concurrent requests. Unfinished requests are lowest for one and highest for four interaction requests for all interaction concepts. It is highest for the trajectory interaction concept.}
\end{figure}

The ART found a significant main effect of \interaction (\F{2}{38}{46.97}, \pminor{0.001}), of \numberOfRequests (\F{3}{57}{37.34}, \pminor{0.001}), and a significant interaction effect of \interaction $\times$ \numberOfRequests on number of missed requests (\F{6}{114}{9.09}, \pminor{0.001}; see \autoref{fig:missingRequests}). While the number of missed requests naturally increased with more \numberOfRequests, the difference between the \interaction increases drastically between \trajectory when four requests are present. This indicates that the effort for \trajectory increases faster with more requests.

\subsection{Lane Deviation}
Lane deviation is a key metric for evaluating path quality during the RO of AVs, as it reflects how well the operator’s inputs align with the optimal path, which is in the middle of a lane. Every lane switch is somewhat risky. It also measures operator performance, highlighting their ability to maintain situational awareness and provide effective guidance. Since the AV handles safety-critical functions like collision avoidance, lane deviation becomes a proxy for potential operational risks and system usability. 

\begin{figure*}[ht!]
\centering
    \begin{subfigure}[c]{0.49\linewidth}
        \includegraphics[width=\linewidth]{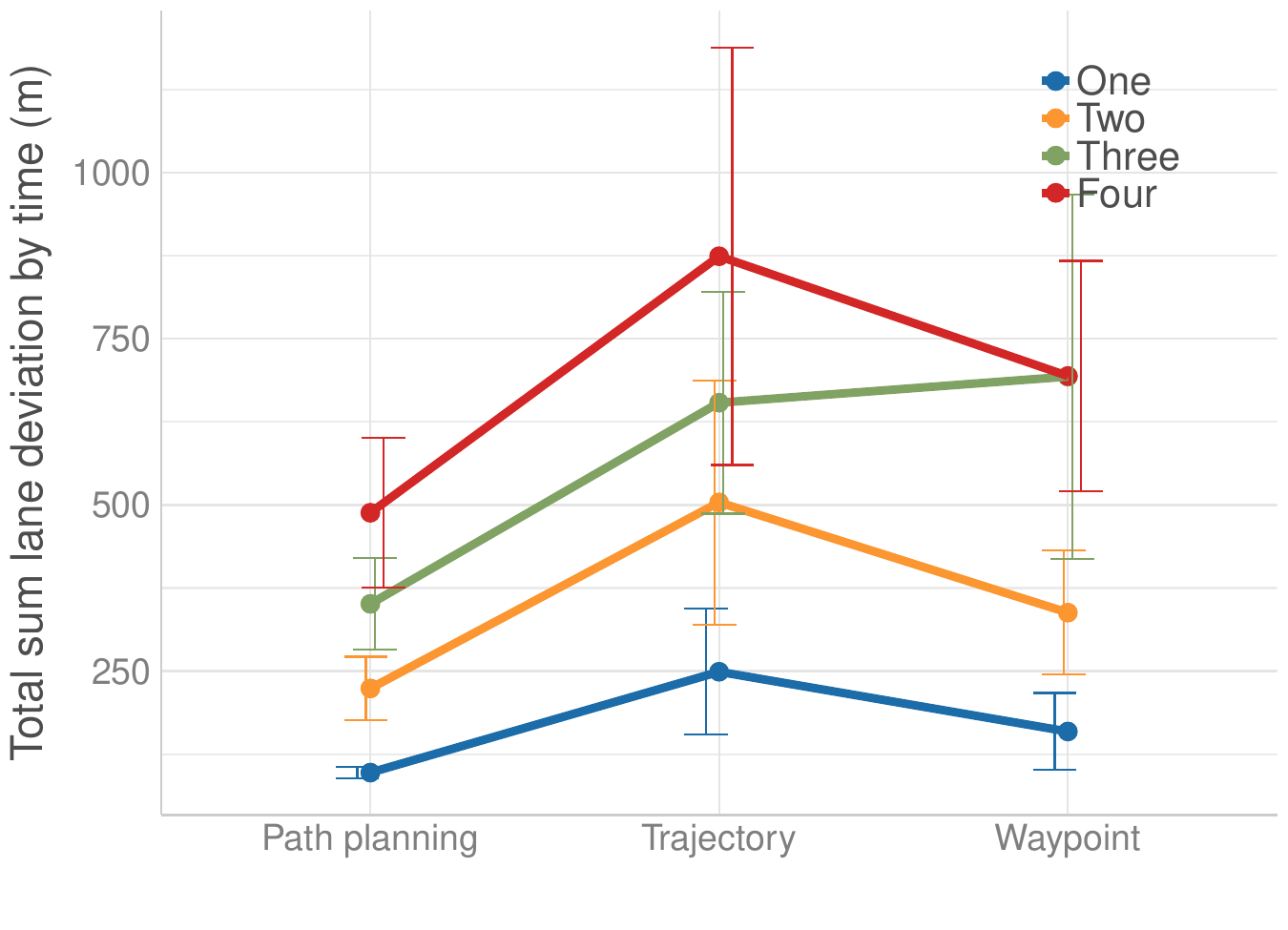}
        \caption{Total sum of time-dependent lane deviation of all active requests.}\label{fig:laneDeviationGraph}
    \end{subfigure}
    \begin{subfigure}[c]{0.49\linewidth}
        \includegraphics[width=\linewidth]{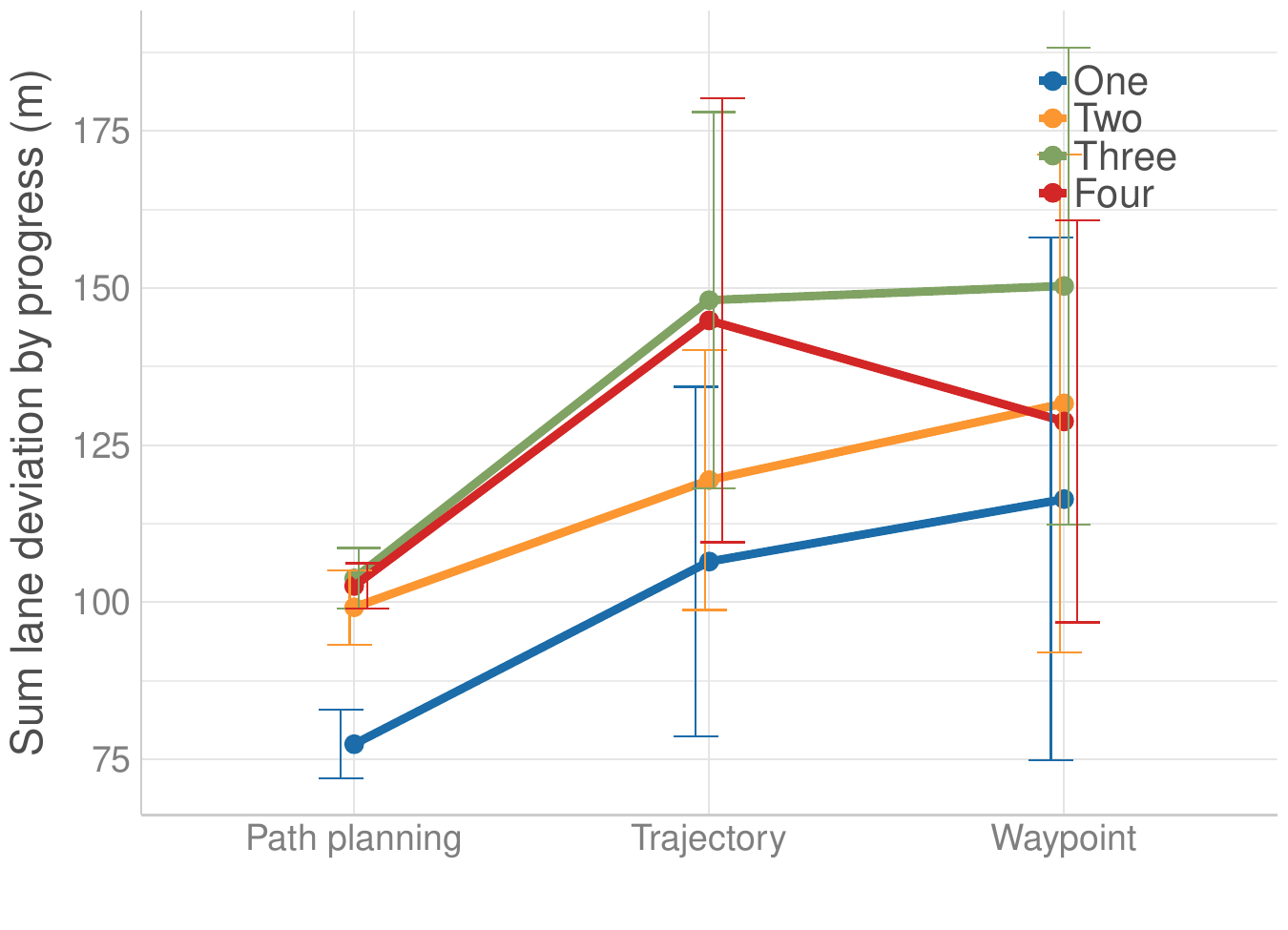}
        \caption{Sum of average lane deviation per request, weighted by progress}\label{fig:laneDeviationGraphfiltered}
    \end{subfigure}
    \caption{Effects on AV lane deviation.}
    \label{fig:laneDeviationGraphs}
    \Description{Effects on vehicle lane deviation. For the lane deviation by time in meters, the highest ones is for four interaction requests, the lowest for one. }
\end{figure*}

\textbf{Time-dependent lane deviation} (see \autoref{fig:laneDeviationGraph}): 
This value is calculated by summing up the current lane deviation (distance to the next lane middle) for every active request in the condition, 10 times per second. The distribution of these values for every participant is visualized in \autoref{fig:laneDeviationGraph}, with the averages for each condition highlighted (lower is better).

The ART found a significant main effect of \interaction (\F{2}{44}{28.89}, \pminor{0.001}) and of \numberOfRequests (\F{3}{66}{82.07}, \pminor{0.001}) on \sumCurrentLaneDeviation. 

A post-hoc test found that Four was significantly higher  (\m{685.56}, \sd{519.48}) in terms of \sumCurrentLaneDeviation compared to Three (\m{566.05}, \sd{458.19}; \padj{0.032}), compared to Two (\m{355.21}, \sd{301.27}; \padjminor{0.001}), compared to One (\m{168.39}, \sd{159.37}; \padjminor{0.001}).
A post-hoc test found that Three was significantly higher (\m{566.05}, \sd{458.19}) in terms of \sumCurrentLaneDeviation compared to Two (\m{355.21}, \sd{301.27}; \padjminor{0.001}) and compared to One (\m{168.39}, \sd{159.37}; \padjminor{0.001}). 
A post-hoc test found that Two was significantly higher  (\m{355.21}, \sd{301.27}) in terms of \sumCurrentLaneDeviation compared to One (\m{168.39}, \sd{159.37}; \padjminor{0.001}).

A post-hoc test found that Trajectory (\m{570.26}, \sd{520.80}; \padjminor{0.001}) and Waypoint (\m{471.05}, \sd{453.22}; \padj{0.006}) were significantly higher in terms of \sumCurrentLaneDeviation compared to Path planning (\m{290.10}, \sd{216.81}). 
A post-hoc test found that Trajectory was significantly higher (\m{570.26}, \sd{520.80}) in terms of \sumCurrentLaneDeviation compared to Waypoint (\m{471.05}, \sd{453.22}; \padj{0.047}).

\textbf{Progress-dependent lane deviation} (see \autoref{fig:laneDeviationGraphfiltered}): 
This value is calculated by first computing each request separately. For every request, all records of lane deviation are mapped to and averaged over the respective current progress in meters (floored to integers). Subsequently, these averages are joined within conditions and participants based on the respective progress value and averaged again. Finally, these averages are summed up over progress, resulting in the sum of the average lane deviation per condition and participant weighted by request progress. The distribution of these values for every participant is visualized in \autoref{fig:laneDeviationGraphfiltered}, with the averages for each condition highlighted (lower is better).

The ART found a significant main effect of \interaction (\F{2}{44}{6.60}, \p{0.003}) and of \numberOfRequests (\F{3}{66}{30.63}, \pminor{0.001}) on \sumCurrentLaneDeviationAv.

A post-hoc test found that Four (\m{125.40}, \sd{65.25}; \padjminor{0.001}),  Three (\m{134.07}, \sd{67.46}; \padjminor{0.001}) and that Two was significantly higher (\m{116.76}, \sd{60.83}; \padjminor{0.001}) in terms of \sumCurrentLaneDeviationAv compared to One (\m{100.09}, \sd{68.29}).

A post-hoc test found that Trajectory (\m{129.71}, \sd{68.05}) and Waypoint (\m{131.79}, \sd{87.19}) were significantly higher  in terms of \sumCurrentLaneDeviationAv compared to Path planning (\m{95.74}, \sd{15.72}; both \padjminor{0.001}).

\begin{comment}
\subsection{Neglected Time}

\begin{figure}
    \centering
    \includegraphics[width=.45\textwidth]{figures/avgMax_interaction.pdf}
    \caption{Interaction effect on neglected time.}
    \label{fig:avgMax_interaction}
    \Description{A graph of the average neglected time of a request, split between the three interaction concepts. For each concept, this is evaluated separately for 1 to 4 total concurrent requests.}
\end{figure}

Neglected time was calculated using the durations of occurrences where requests were neglected and averaging them.

The ART found a significant main effect of \interaction (\F{2}{44}{11.27}, \pminor{0.001}), of \numberOfRequests (\F{3}{66}{131.54}, \pminor{0.001}), and a significant interaction effect of \interaction $\times$ \numberOfRequests on neglected time (\F{6}{132}{4.57}, \pminor{0.001}; see \autoref{fig:avgMax_interaction}). While neglected time was almost zero for all \interaction when one request was issued, for two requests, \trajectory was lowest, but \trajectory was highest with three or four requests.

\end{comment}

\subsection{Neglected Time}

\begin{figure}
    \centering
    \includegraphics[width=.45\textwidth]{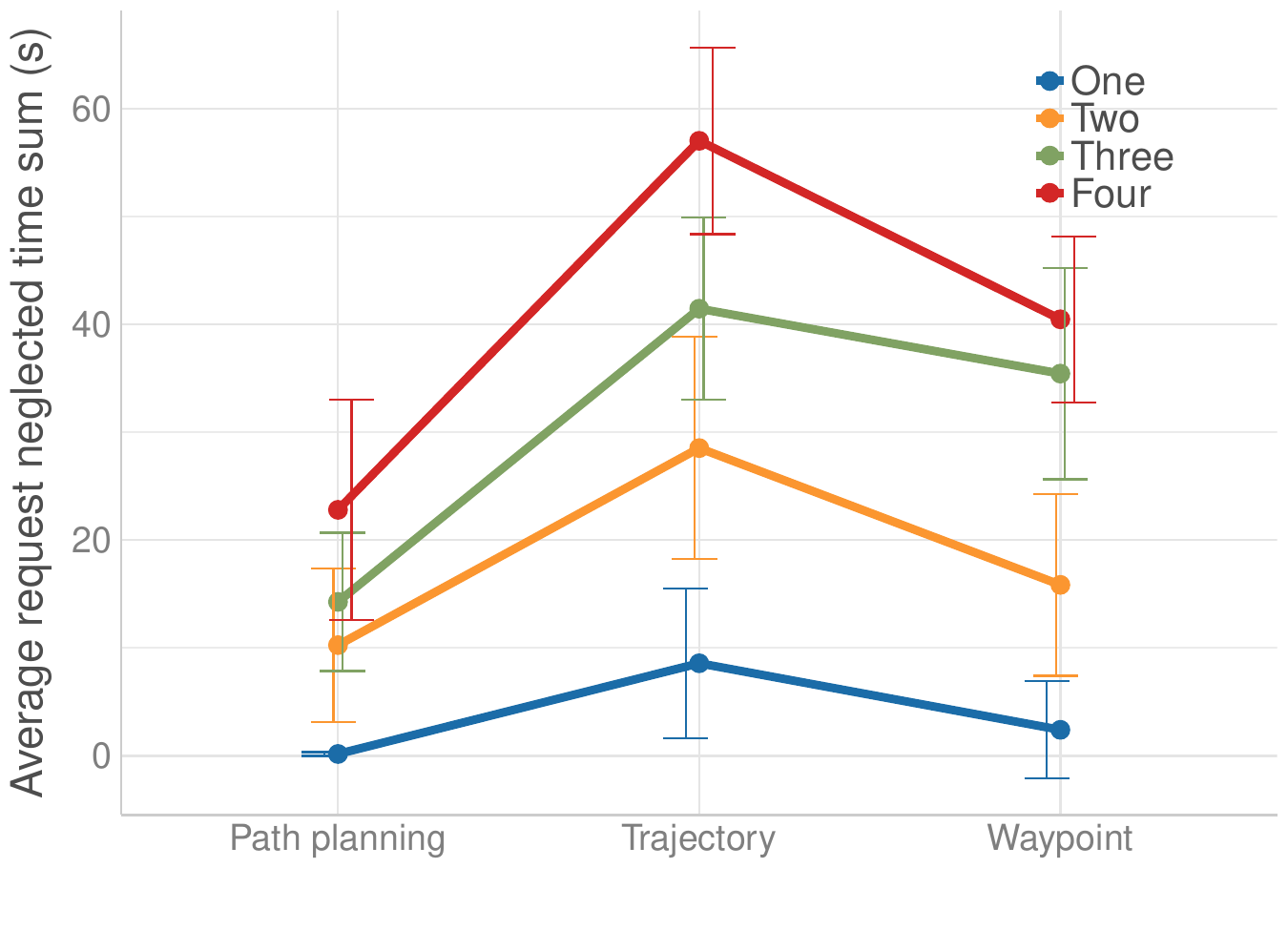}
    \caption{Interaction effect on neglected time.}
    \label{fig:avgMax_interaction}
    \Description{A graph of the average neglected time of a request split between the three interaction concepts. For each concept, this is evaluated separately for 1 to 4 total concurrent requests. The neglected time sum increases with the number of requests. the difference is low for path planning and highest for trajectory.}
\end{figure}

Neglected time was calculated by adding up the durations during which requests were neglected (i.e., a request is issued but not worked on in the main windows) and then calculating the average of this sum (i.e., cumulative neglected time total divided by neglect counts). A request is considered neglected when the RO is required to take action to enable a stopped vehicle to resume driving but does not as soon as a request is emitted.

The ART found a significant main effect of \interaction (\F{2}{44}{60.38}, \pminor{0.001}), of \numberOfRequests (\F{3}{66}{89.67}, \pminor{0.001}), and a significant interaction effect of \interaction $\times$ \numberOfRequests on neglected time (\F{6}{132}{8.86}, \pminor{0.001}; see \autoref{fig:avgMax_interaction})

The average neglected time per request naturally increased with more \numberOfRequests. At each level of \numberOfRequests, \pathPlanning incurs the least amount of neglect time, while \trajectory incurs the highest.

\subsection{Mouse Movements}

\begin{figure}
    \centering
    \includegraphics[width=.5\textwidth]{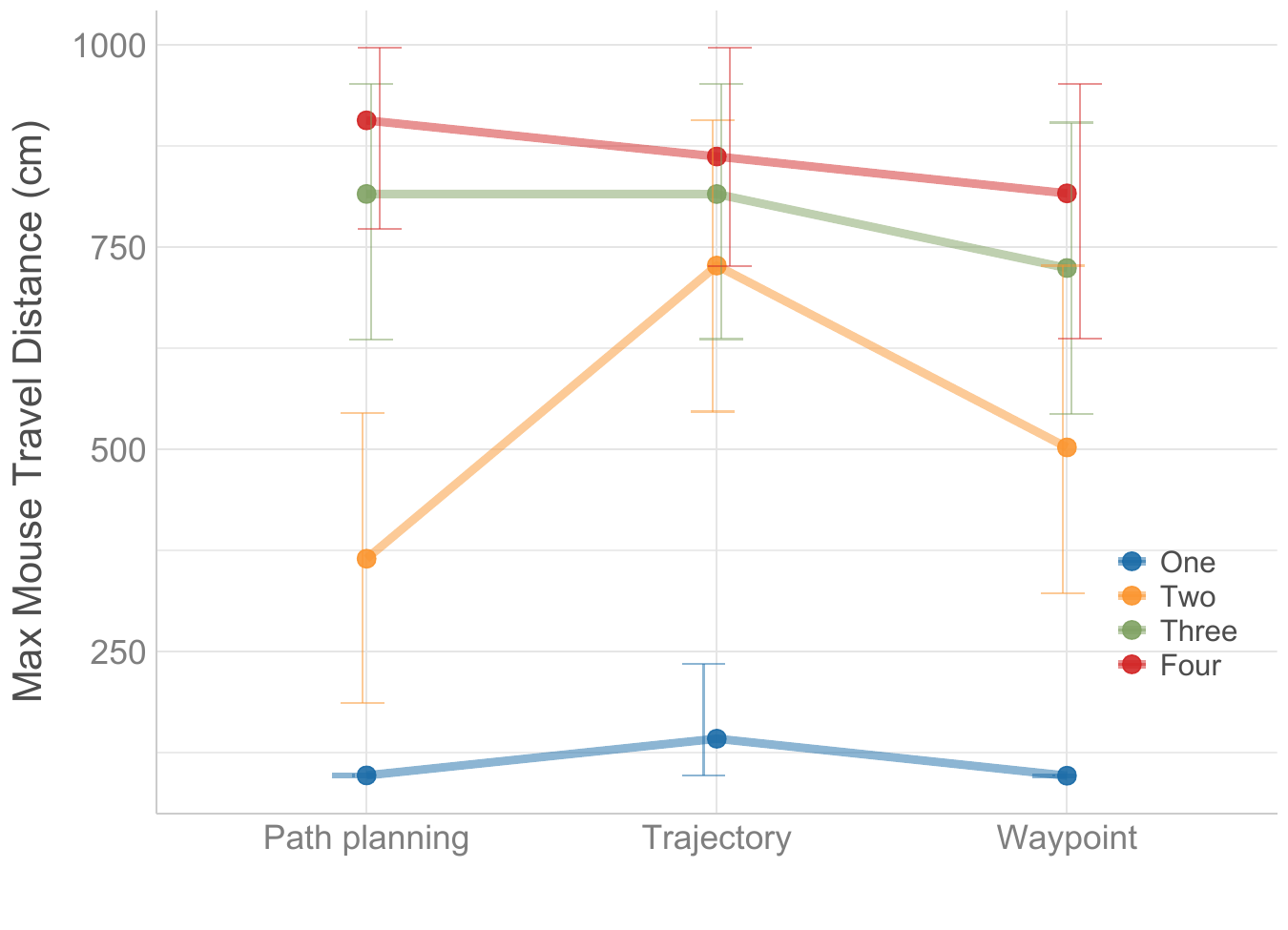}
    \caption{Interaction effect on mouse travel distance (cm).}
    \label{fig:maxMouseTravelDistanceInCm_interaction}
    \Description{A graph of the total mouse travel distance split between the three interaction concepts. For each concept, this is evaluated separately for 1 to 4 total concurrent requests. For one request, the distance is the lowest. For two or more requests, the distance is far bigger. For two requests, it is biggest with trajectory, with path planning being considerably lower. }
\end{figure}

The ART found a significant main effect of \interaction (\F{2}{38}{7.30}, \p{0.002}), of \numberOfRequests (\F{3}{57}{65.80}, \pminor{0.001}), and a significant interaction effect of \interaction $\times$ \numberOfRequests on mouse travel distance (\F{6}{114}{5.41}, \pminor{0.001}; see \autoref{fig:maxMouseTravelDistanceInCm_interaction}). While the order (from least to most movements) was consistently higher, the more requests were involved, for the \trajectory, already two requests led to very high mouse movement.

%\subsection{Transitions}

%\begin{figure}
%    \centering
%    \includegraphics[width=.75\linewidth]{figures/Graph_interaction_Path planning_numberOfRequests_Two.pdf}
%    \caption{Transition graph for \pathPlanning and \numberOfRequests two.}
%    \label{fig:Graph_interaction_Path}
%\end{figure}

%\autoref{fig:Graph_interaction_Path}
%\todo{this is one of 12 graphs - we would have to describe them in detail-  could be appendix}

\subsection{Task Load}
\begin{figure*}[ht!]
  \centering
  \begin{subfigure}{0.33\textwidth}
    \centering
    \includegraphics[width=\textwidth]{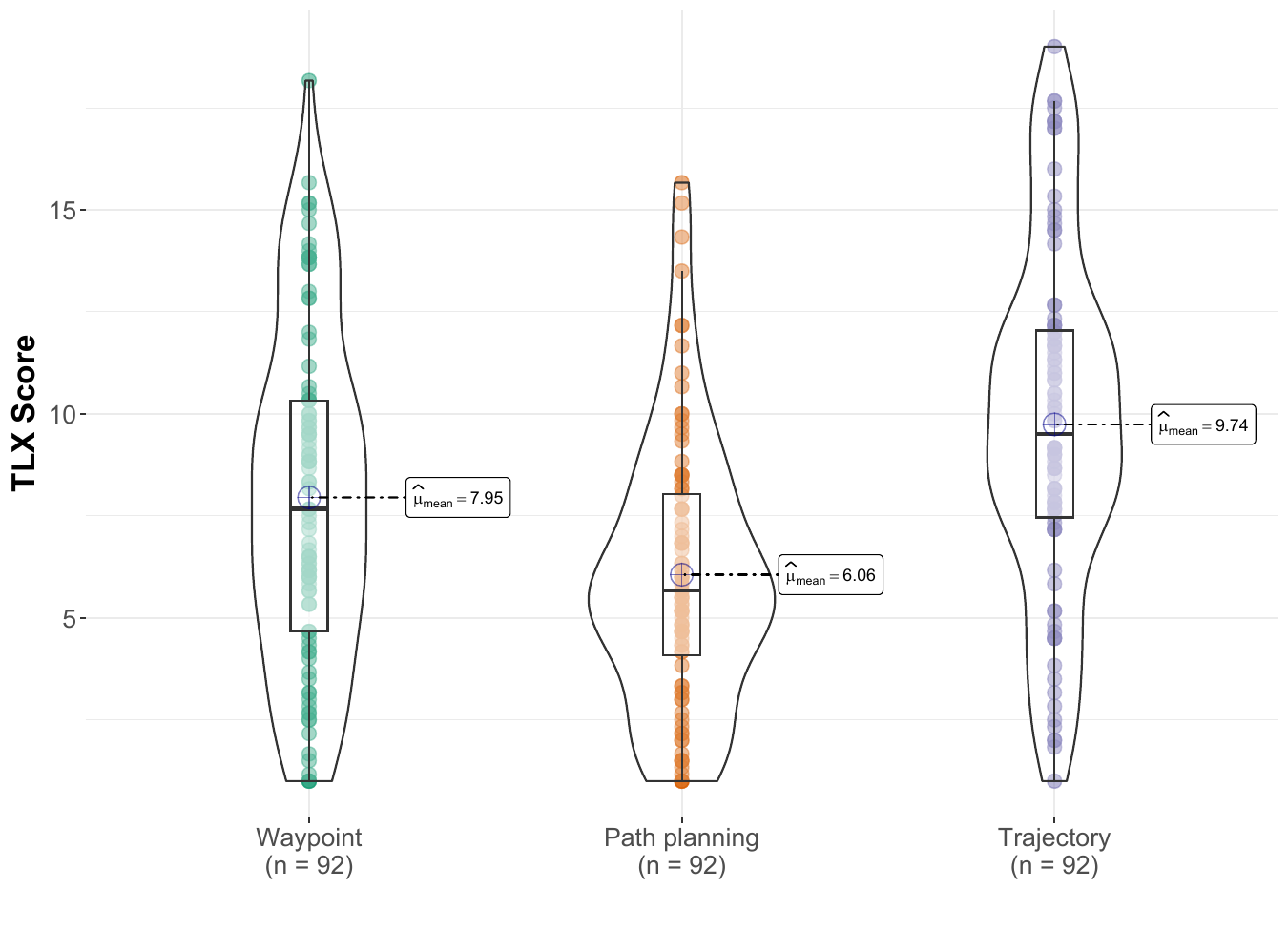}
    \caption{TLX Score per \interaction.}
    \label{fig:interaction_tlx}
  \end{subfigure}%
  \begin{subfigure}{0.33\textwidth}
    \centering
    \includegraphics[width=\textwidth]{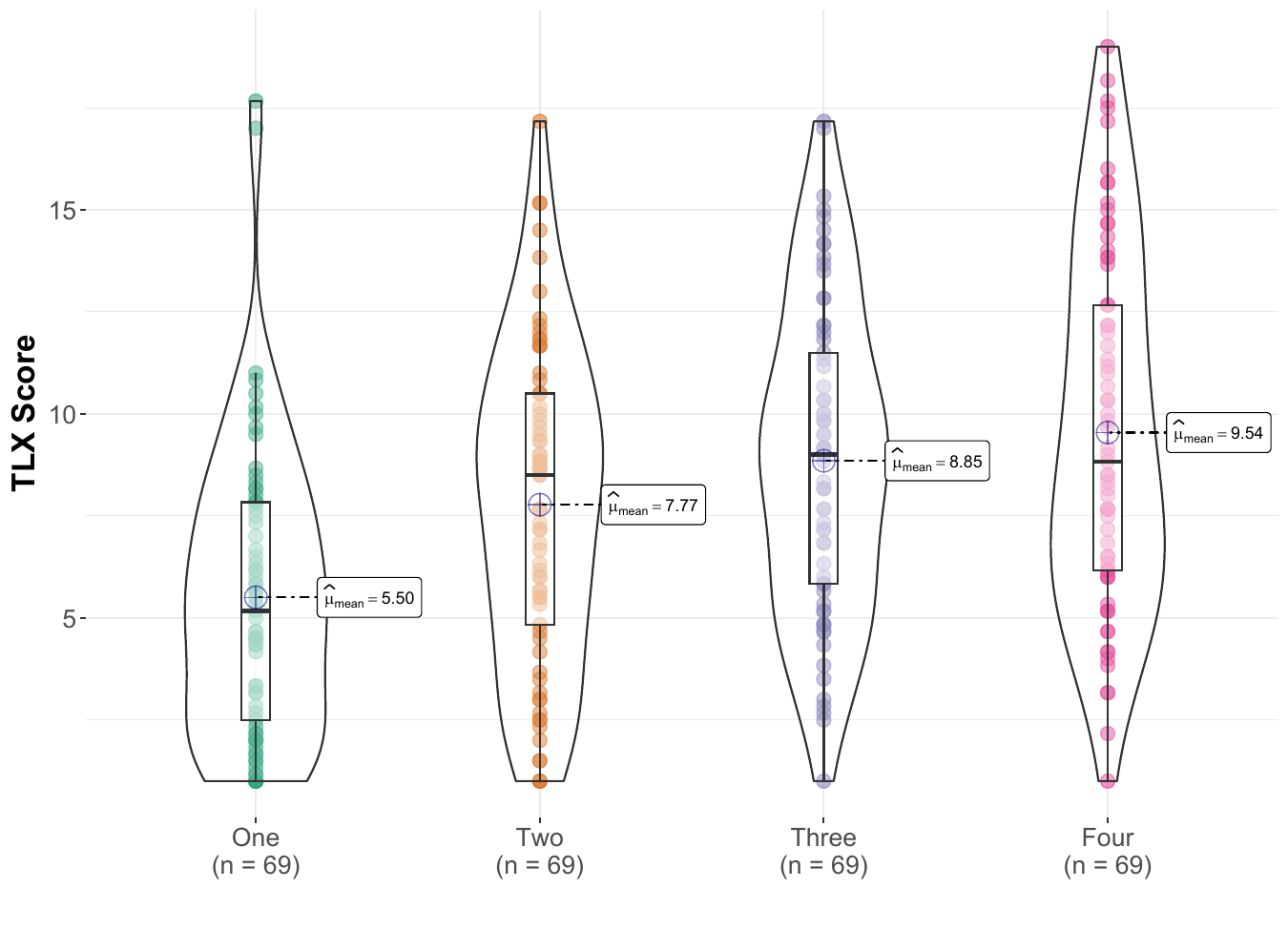}
    \caption{TLX Score per \numberOfRequests.}
    \label{fig:nr_tlx}
  \end{subfigure}
    \caption{TLX Score per \interaction and \numberOfRequests.}
    \label{fig:tlx}
    \Description{TLX Score per interaction concept and number of requests.}
\end{figure*}

The ART found a significant main effect of \interaction (\F{2}{44}{20.83}, \pminor{0.001}; see \autoref{fig:interaction_tlx}) and of \numberOfRequests on TLX score (\F{3}{66}{22.02}, \pminor{0.001}; see \autoref{fig:nr_tlx}).

A post-hoc test found that Trajectory (\m{9.74}, \sd{4.26}) and Waypoint (\m{7.95}, \sd{4.11}) were significantly worse in terms of \tlxScore compared to \pathPlanning (\m{6.06}, \sd{3.30}; \padjminor{0.001}) and that Trajectory was significantly worse  (\m{9.74}, \sd{4.26}) than Waypoint (\m{7.95}, \sd{4.11}; \padj{0.005}). 

A post-hoc test found that Four  (\m{9.54}, \sd{4.41}; \padjminor{0.00}), Three (\m{8.85}, \sd{3.81}; \padjminor{0.00}), and Two requests were significantly worse  (\m{7.77}, \sd{3.86}; \padj{0.002}) in terms of \tlxScore than One request (\m{5.50}, \sd{3.51}).

\subsection{System Usability Score}

\begin{figure*}[ht!]
  \centering
  \begin{subfigure}{0.33\textwidth}
    \centering
    \includegraphics[width=\textwidth]{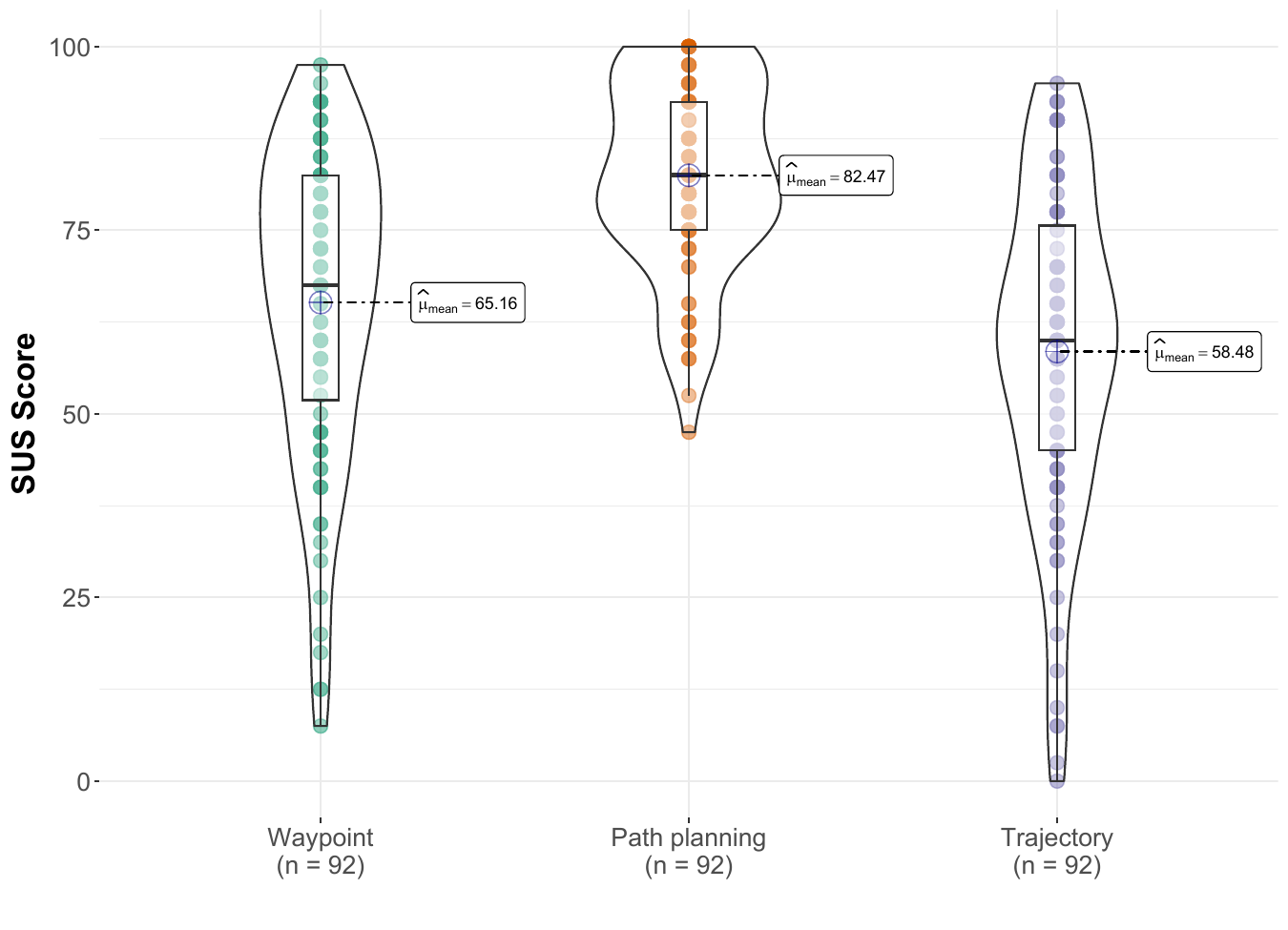}
    \caption{SUS Score per \interaction.}
    \label{fig:interaction_sus}
  \end{subfigure}%
  \begin{subfigure}{0.33\textwidth}
    \centering
    \includegraphics[width=\textwidth]{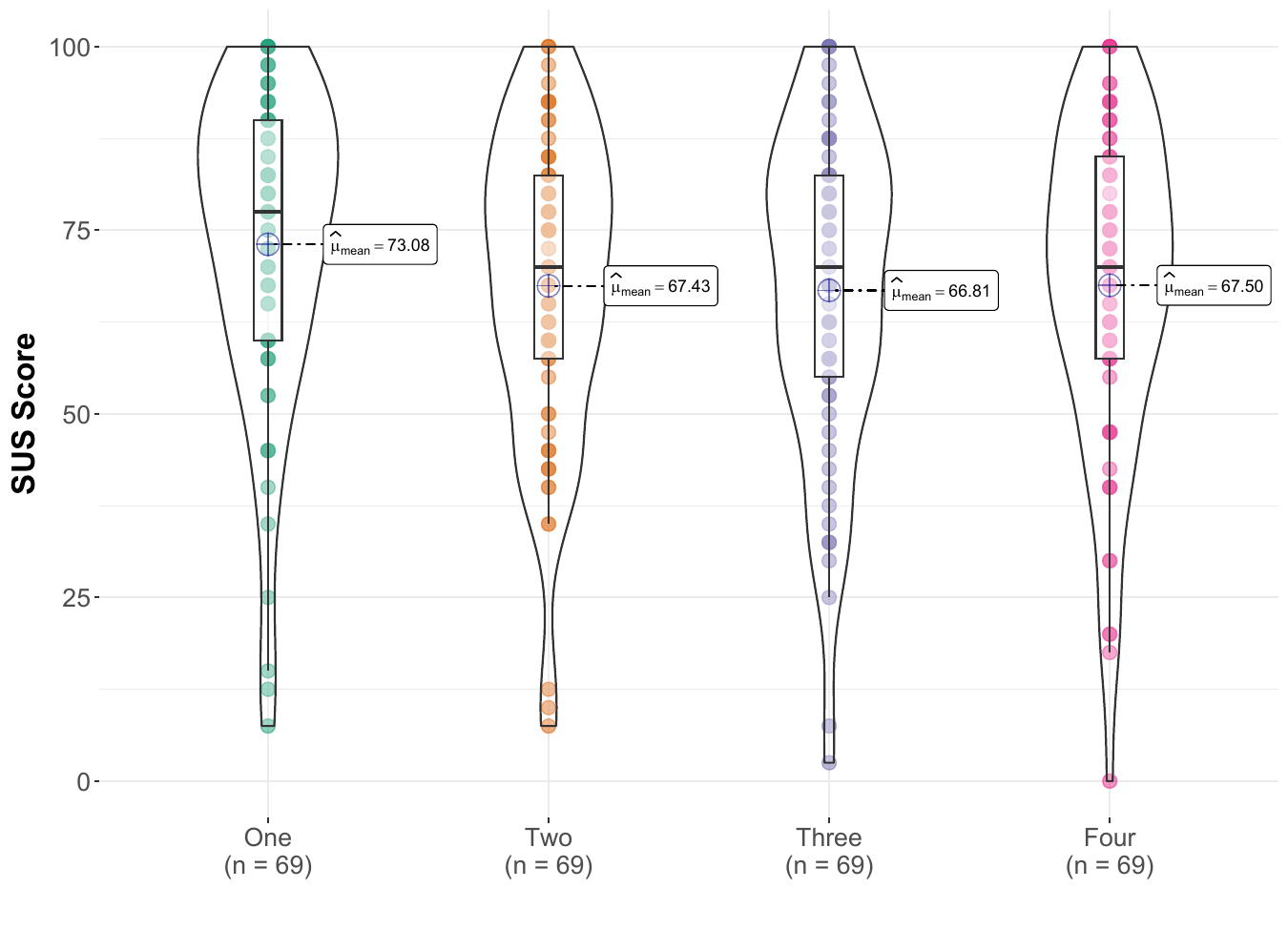}
    \caption{SUS Score per \numberOfRequests.}
    \label{fig:nr_sus}
  \end{subfigure}
    \caption{SUS Score per \interaction and \numberOfRequests.}
    \label{fig:sus}
    \Description{SUS Score per interaction concept and number of requests.}
\end{figure*}

The ART found a significant main effect of \interaction (\F{2}{44}{23.96}, \pminor{0.001}; see \autoref{fig:interaction_sus}) and of \numberOfRequests on the SUS score (\F{3}{66}{4.16}, \p{0.009}; no significant post-hoc results; see \autoref{fig:nr_sus}). 

% interaction
A post-hoc test found that \pathPlanning was significantly higher  (\m{82.47}, \sd{13.06}) in terms of \SUSScore compared to Trajectory (\m{58.48}, \sd{22.45}; \padjminor{0.001}) and Waypoint (\m{65.16}, \sd{21.27}; \padjminor{0.001}). 
A post-hoc test found that Waypoint was significantly higher  (\m{65.16}, \sd{21.27}) in terms of \SUSScore than Trajectory (\m{58.48}, \sd{22.45}; \padj{0.041}). 

% put it in brackets
%Post-hoc tests found no significant effects regarding \numberOfRequests.

\subsection{Usefulness and Satisfying}\label{sec:acceptance}

\begin{figure*}[ht!]
  \centering

  \begin{subfigure}{0.48\textwidth}
    \centering
    \includegraphics[width=\textwidth]{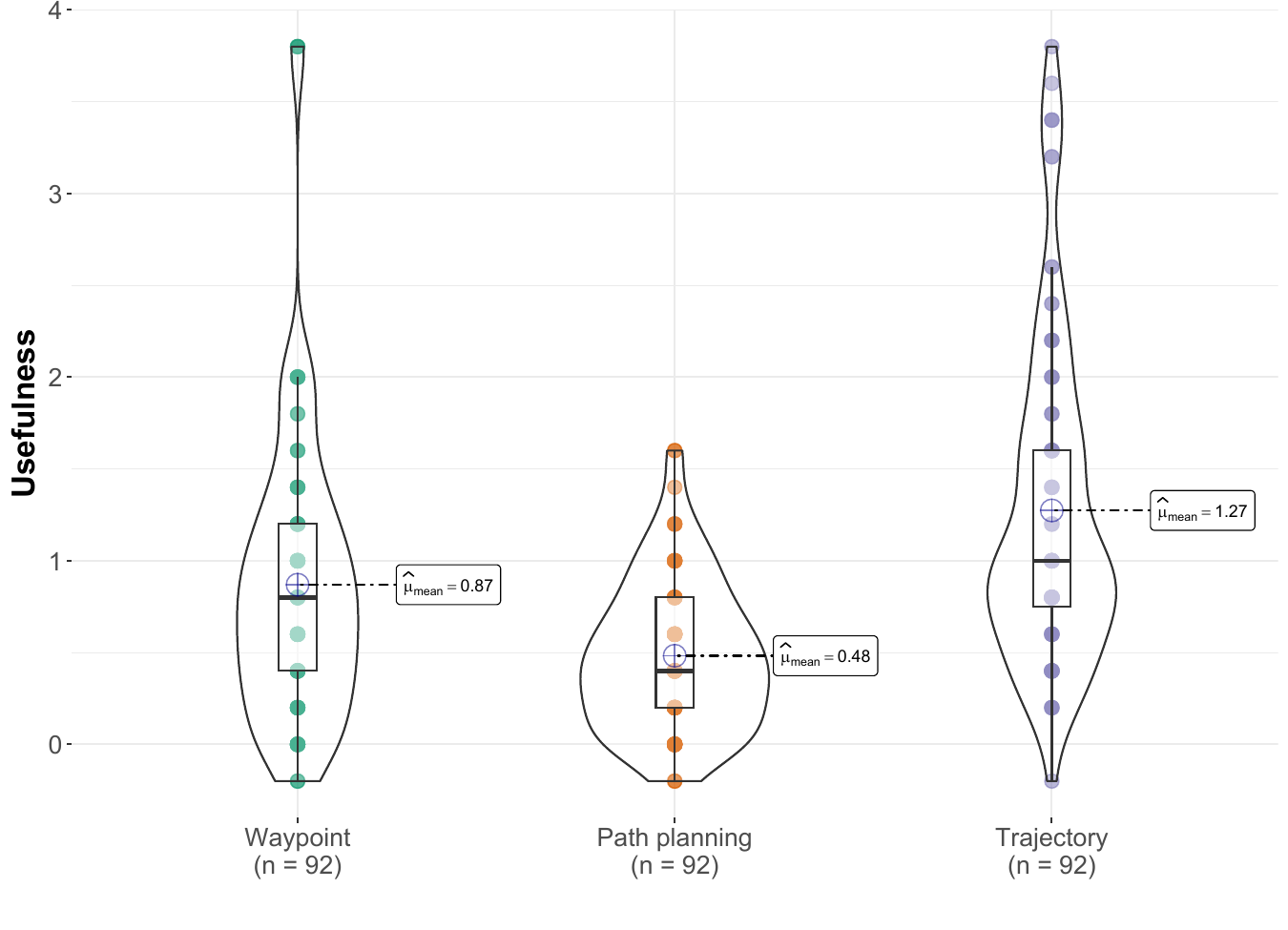}
    \caption{Usefulness per \interaction.}
    \label{fig:interaction_usefulness}
  \end{subfigure}%
  \begin{subfigure}{0.48\textwidth}
    \centering
    \includegraphics[width=\textwidth]{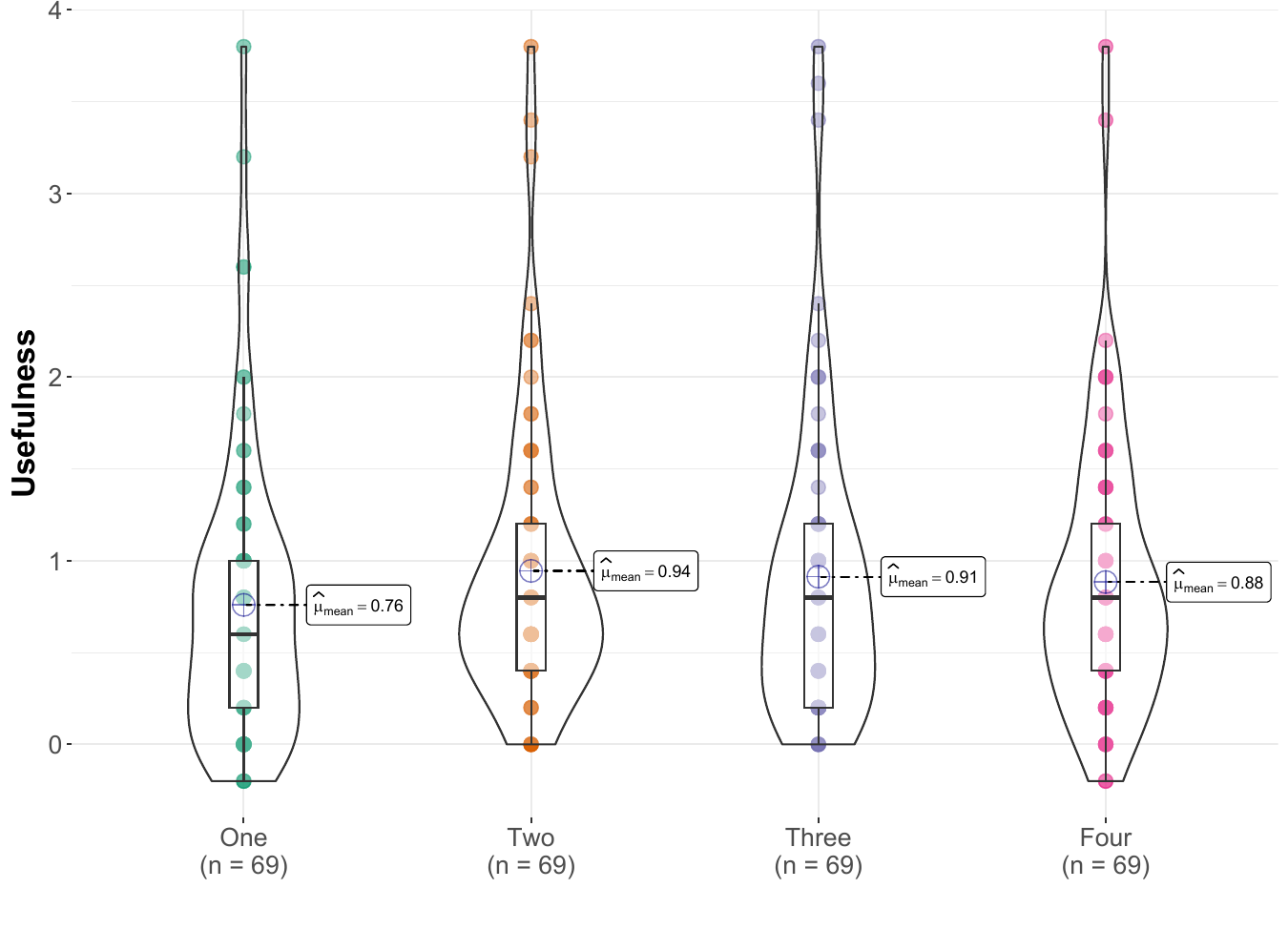}
    \caption{Usefulness per \numberOfRequests.}
    \label{fig:nr_usefulness}
  \end{subfigure}
  \begin{subfigure}{0.48\textwidth}
    \centering
    \includegraphics[width=\textwidth]{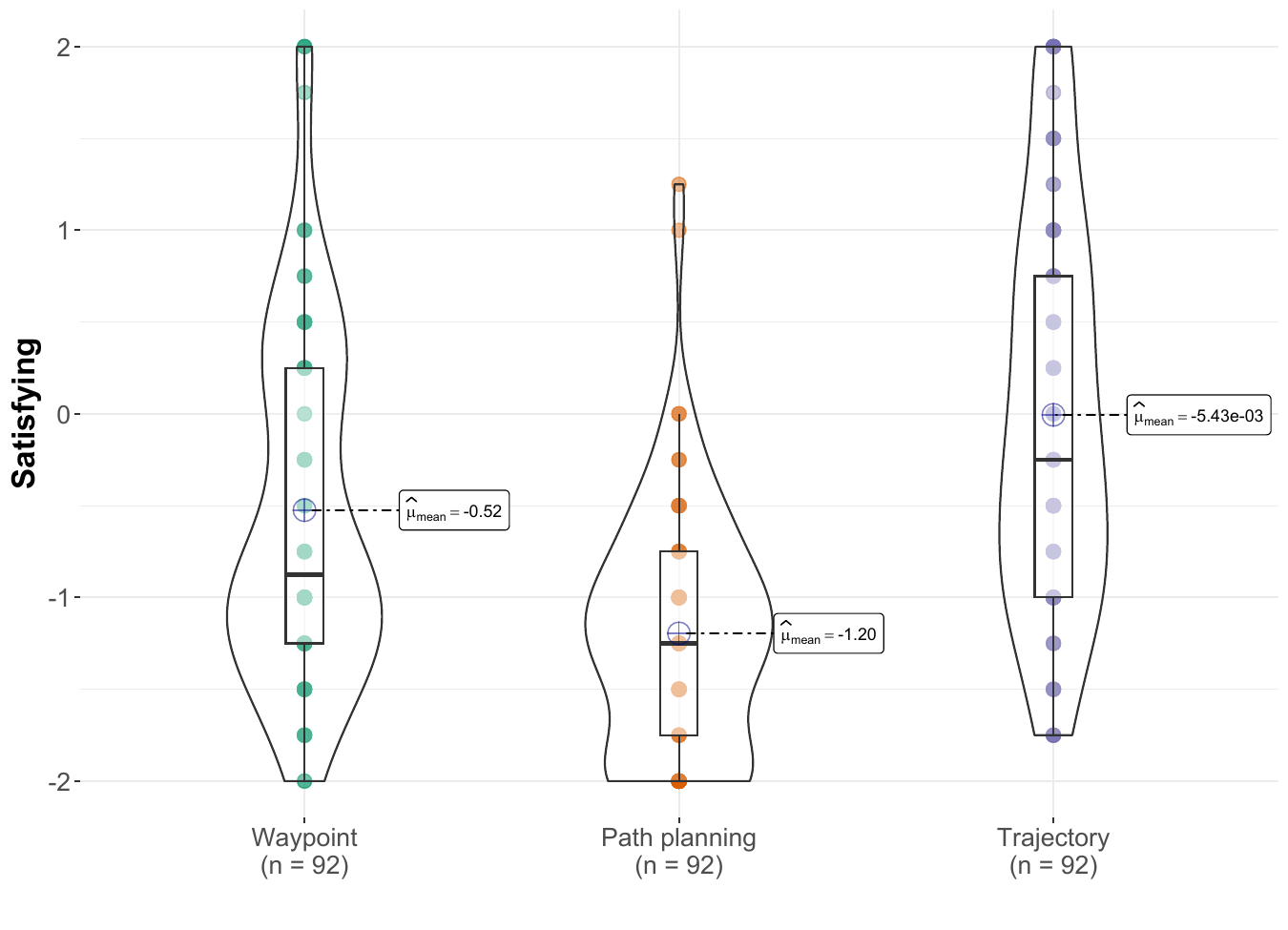}
    \caption{Satisfying per \interaction.}
    \label{fig:interaction_satisfying}
  \end{subfigure}%
  \begin{subfigure}{0.48\textwidth}
    \centering
    \includegraphics[width=\textwidth]{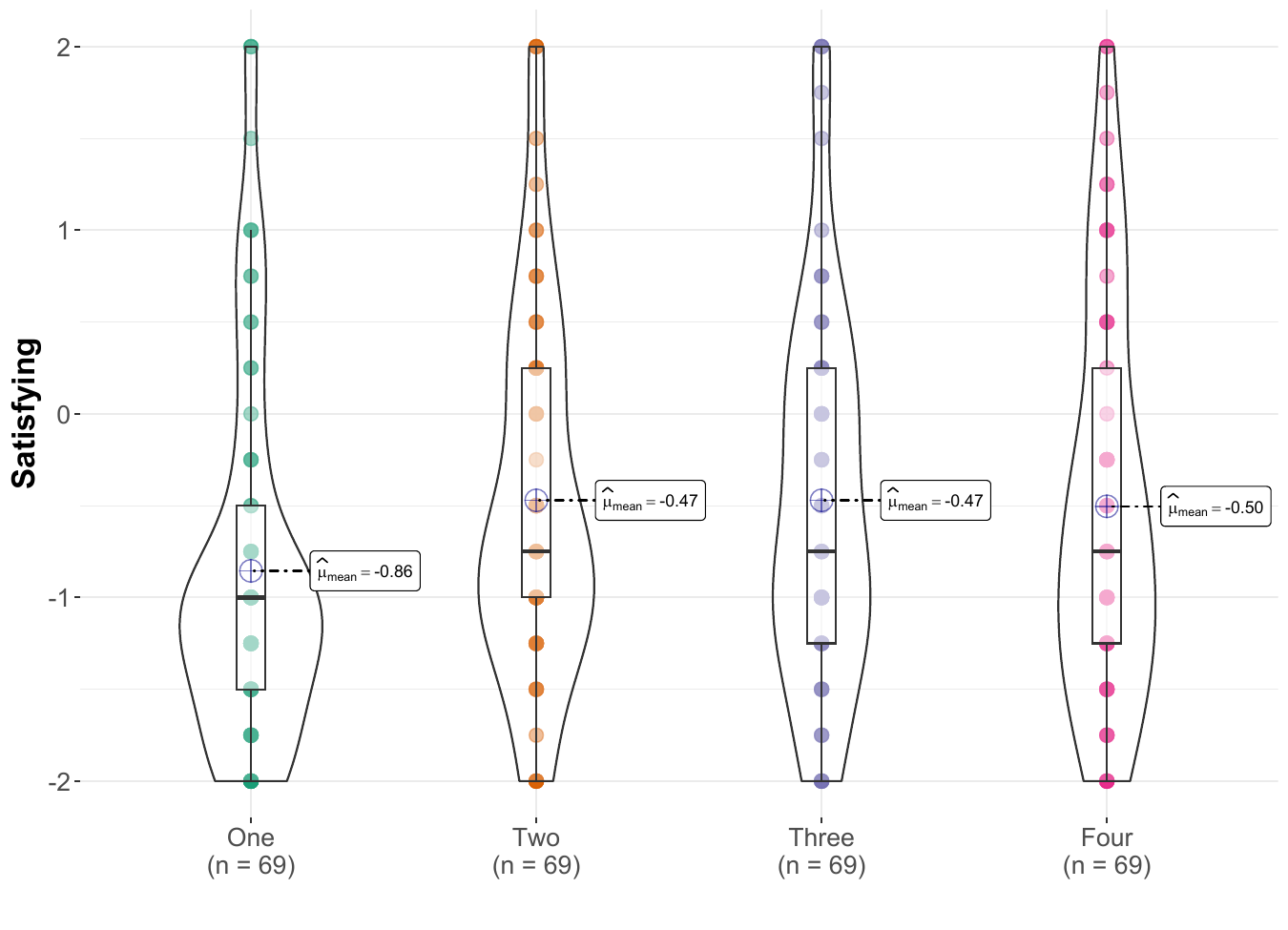}
    \caption{Satisfying per \numberOfRequests.}
    \label{fig:nr_satisfying}
  \end{subfigure}
    \caption{Usefulness and Satisfying per \interaction and \numberOfRequests.}
    \label{fig:usefulness_and_satisfying}
    \Description{Usefulness and Satisfying per interaction concept and number of requests.}
\end{figure*}

The ART found a significant main effect of \interaction (\F{2}{44}{14.60}, \pminor{0.001}) and of \numberOfRequests on Usefulness (\F{3}{66}{4.39}, \p{0.007} (lower is better); see \autoref{fig:usefulness_and_satisfying}). 

A post-hoc test found that Trajectory (\m{1.27}, \sd{0.87}) and Waypoint (\m{0.87}, \sd{0.79}) were significantly worse in terms of \AOAUsefulness compared to \pathPlanning (\m{0.48}, \sd{0.40}; both \padjminor{0.001}) and that Trajectory was significantly worse  (\m{1.27}, \sd{0.87}) compared to Waypoint (\padjminor{0.001}).

%Satisfying
The ART found a significant main effect of \interaction (\F{2}{44}{22.25}, \pminor{0.001}) and of \numberOfRequests on Satisfying (\F{3}{66}{7.01}, \pminor{0.001} (lower is better); see \autoref{fig:usefulness_and_satisfying}).

A post-hoc test found that Trajectory (\m{-0.01}, \sd{1.05}) and Waypoint (\m{-0.52}, \sd{0.95}) were significantly worse in terms of \AOASatisfying compared to \pathPlanning (\m{-1.20}, \sd{0.66}; both \padjminor{0.001}) and that Trajectory was significantly worse  (\m{-0.01}, \sd{1.05}) than Waypoint (\padjminor{0.001}).

A post-hoc test found that Two was significantly higher (\m{-0.47}, \sd{0.99}) in terms of \AOASatisfying compared to One (\m{-0.86}, \sd{0.99}; \padj{0.048}).

\subsection{Movement Patterns}
\begin{figure*}
  \centering

  \begin{subfigure}{0.245\textwidth}
    \centering
    \includegraphics[width=\textwidth]{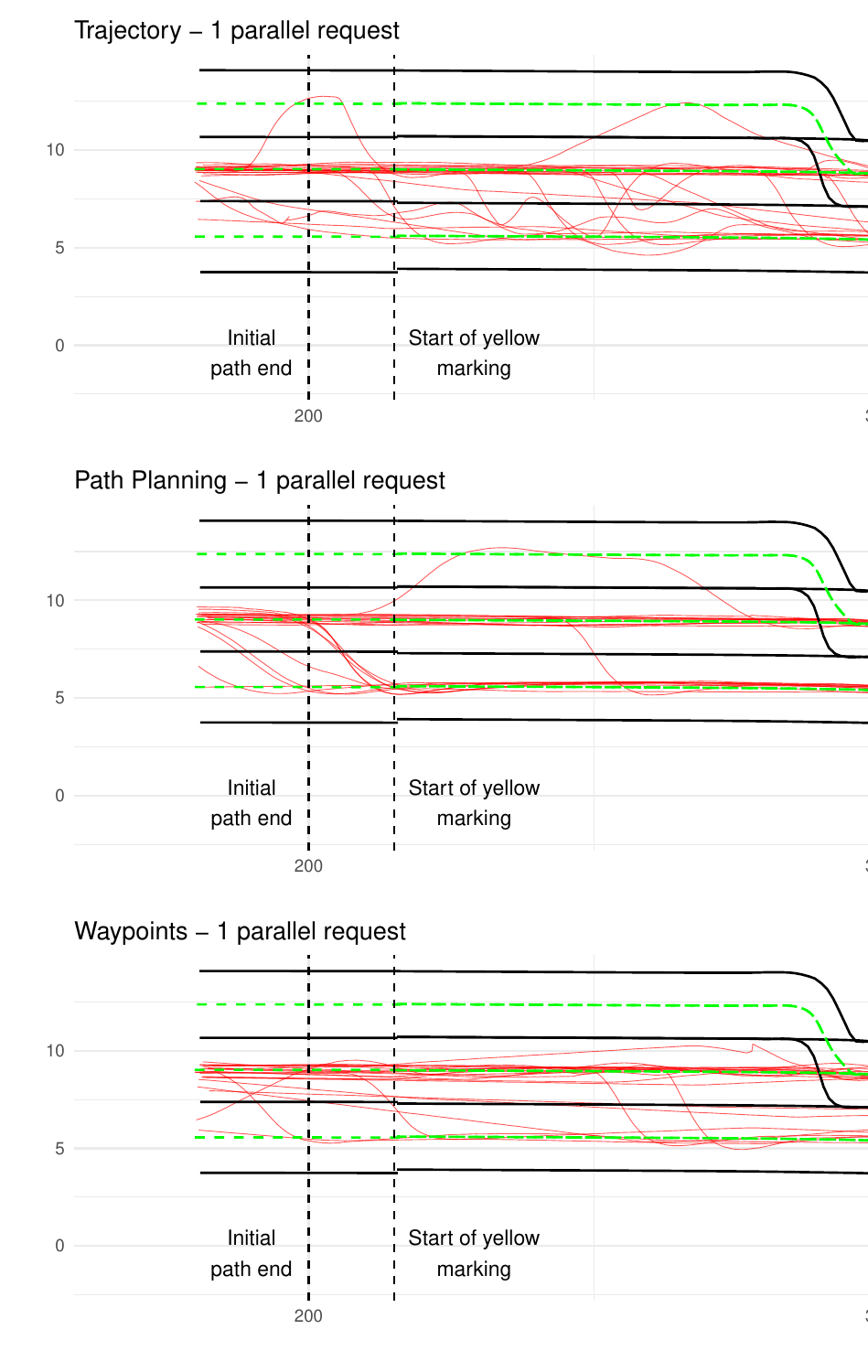}
    \caption{One request.}
    \label{fig:Pattern1-left}
  \end{subfigure}%
  \begin{subfigure}{0.245\textwidth}
    \centering
    \includegraphics[width=\textwidth]{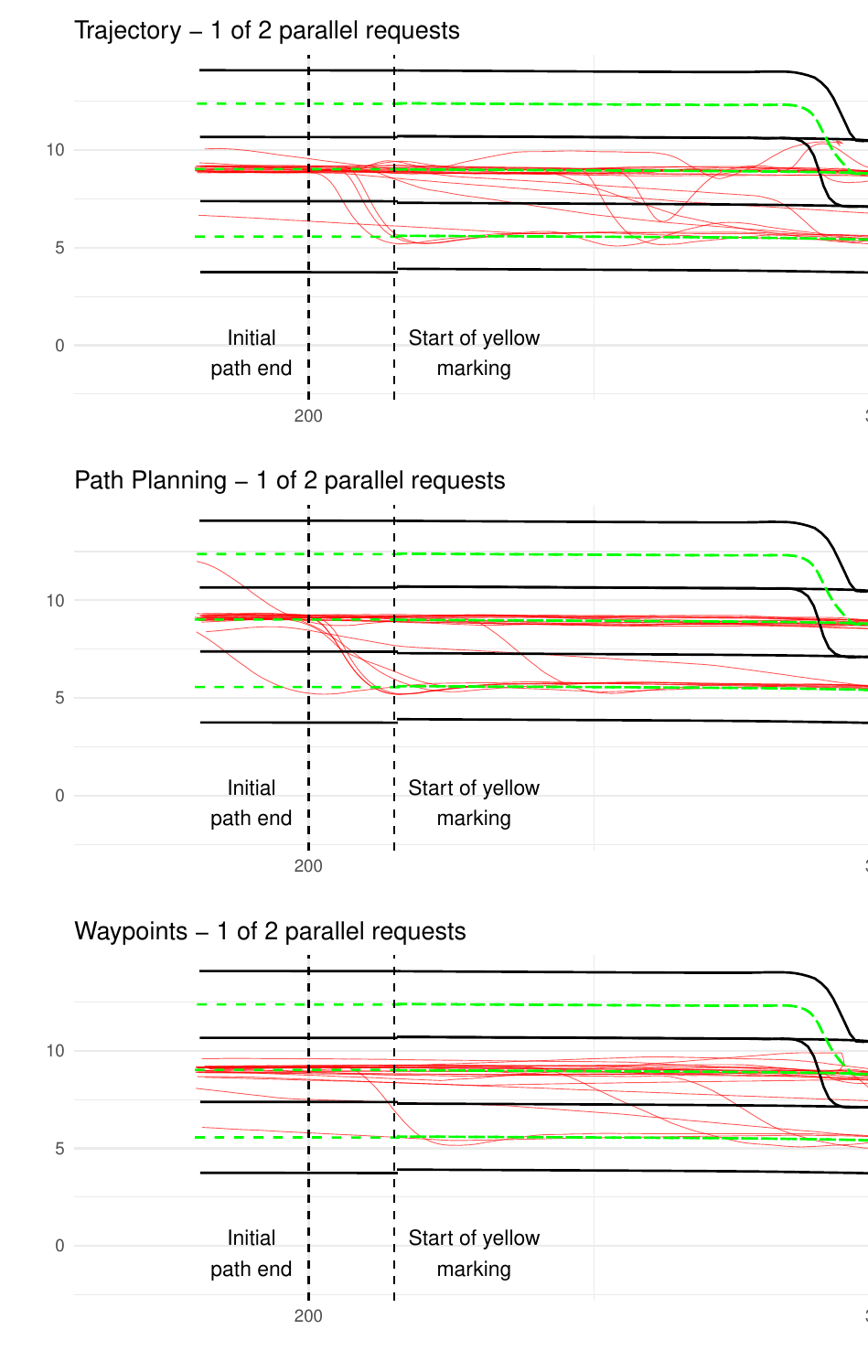}
    \caption{Two requests.}
    \label{fig:Pattern2-left}
  \end{subfigure}%
  \begin{subfigure}{0.245\textwidth}
    \centering
    \includegraphics[width=\textwidth]{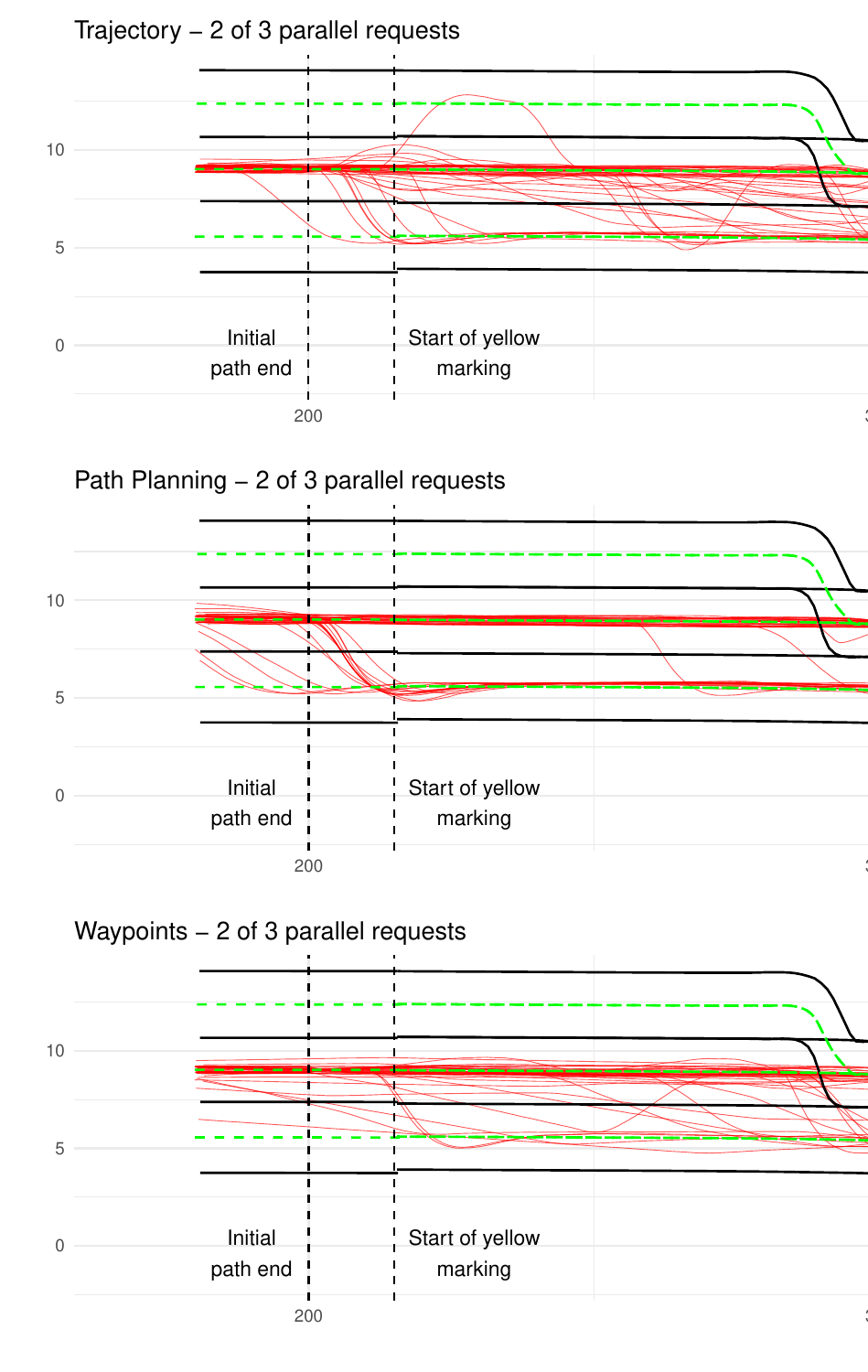}
    \caption{Three requests.}
    \label{fig:Pattern3-left}
  \end{subfigure}%
  \begin{subfigure}{0.245\textwidth}
    \centering
    \includegraphics[width=\textwidth]{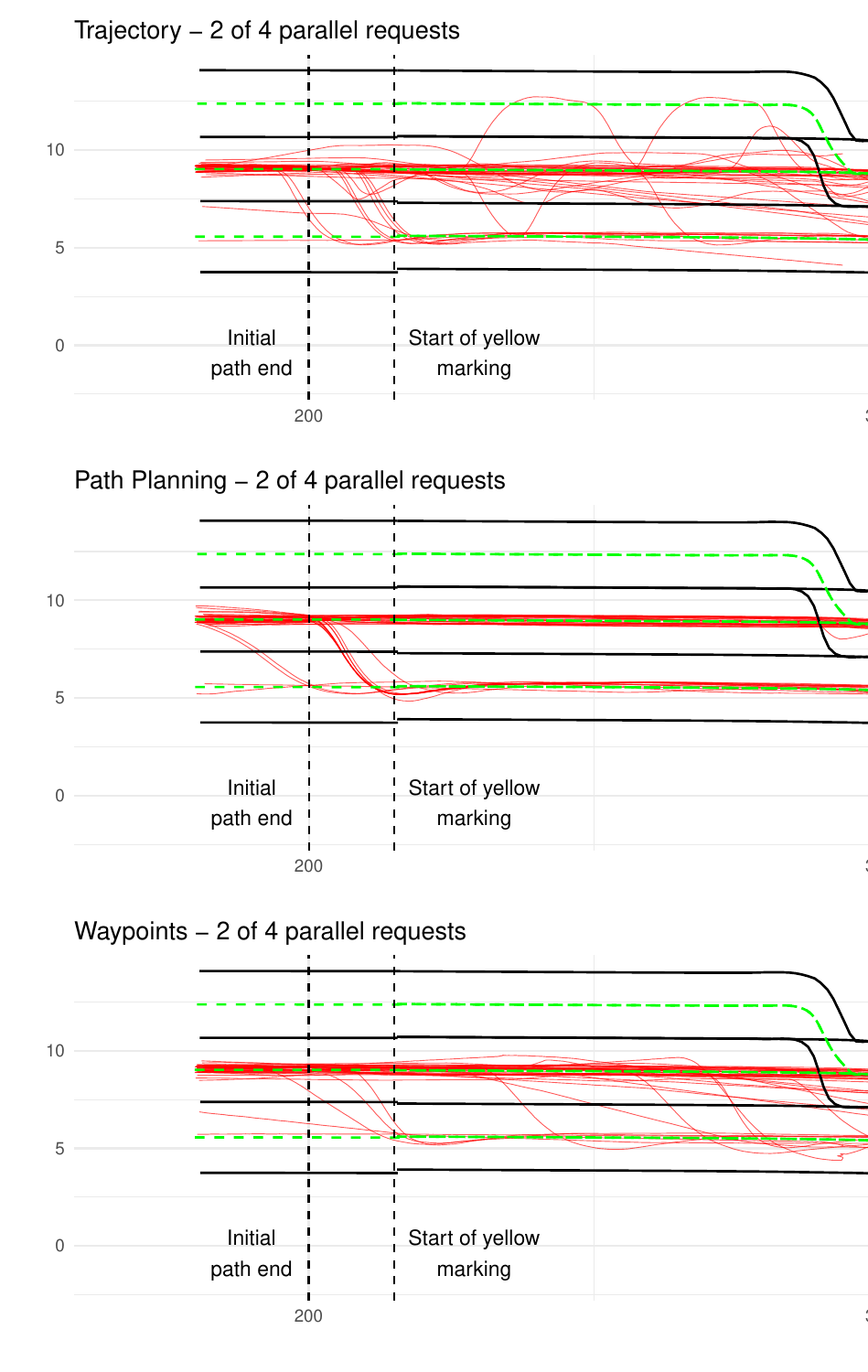}
    \caption{Four requests.}
    \label{fig:Pattern4-left}
  \end{subfigure}
    \caption{Movement patterns when approaching the left-sided road works.}
    \label{fig:startPatterns-left}
    \Description{Movement patterns when approaching the left-sided road works.}
\end{figure*}

\begin{figure*}[ht]
  \centering
  \begin{subfigure}{0.33\textwidth}
    \centering
    \includegraphics[width=\textwidth]{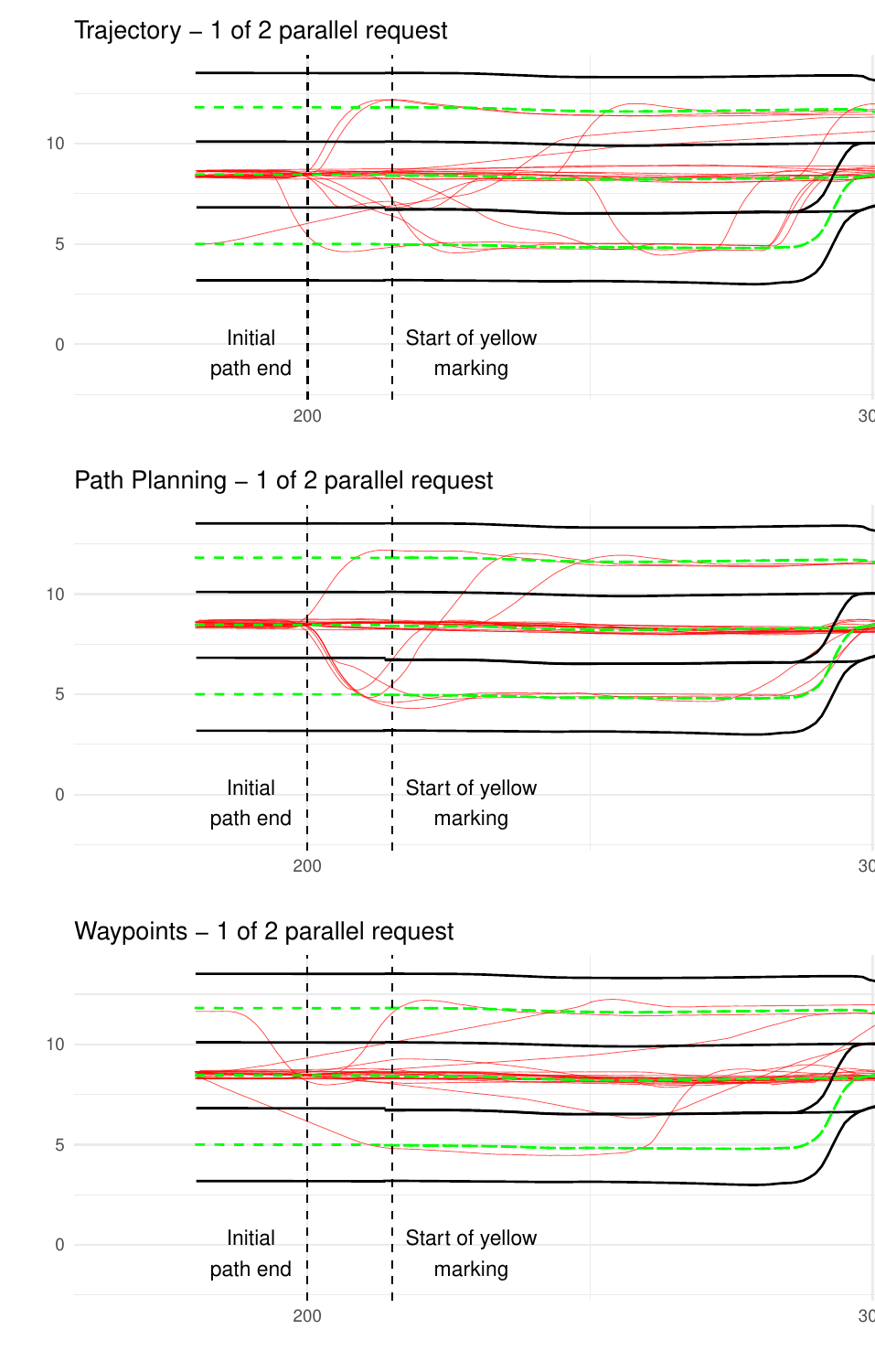}
    \caption{Two requests.}
    \label{fig:Pattern2-right}
  \end{subfigure}%
  \begin{subfigure}{0.33\textwidth}
    \centering
    \includegraphics[width=\textwidth]{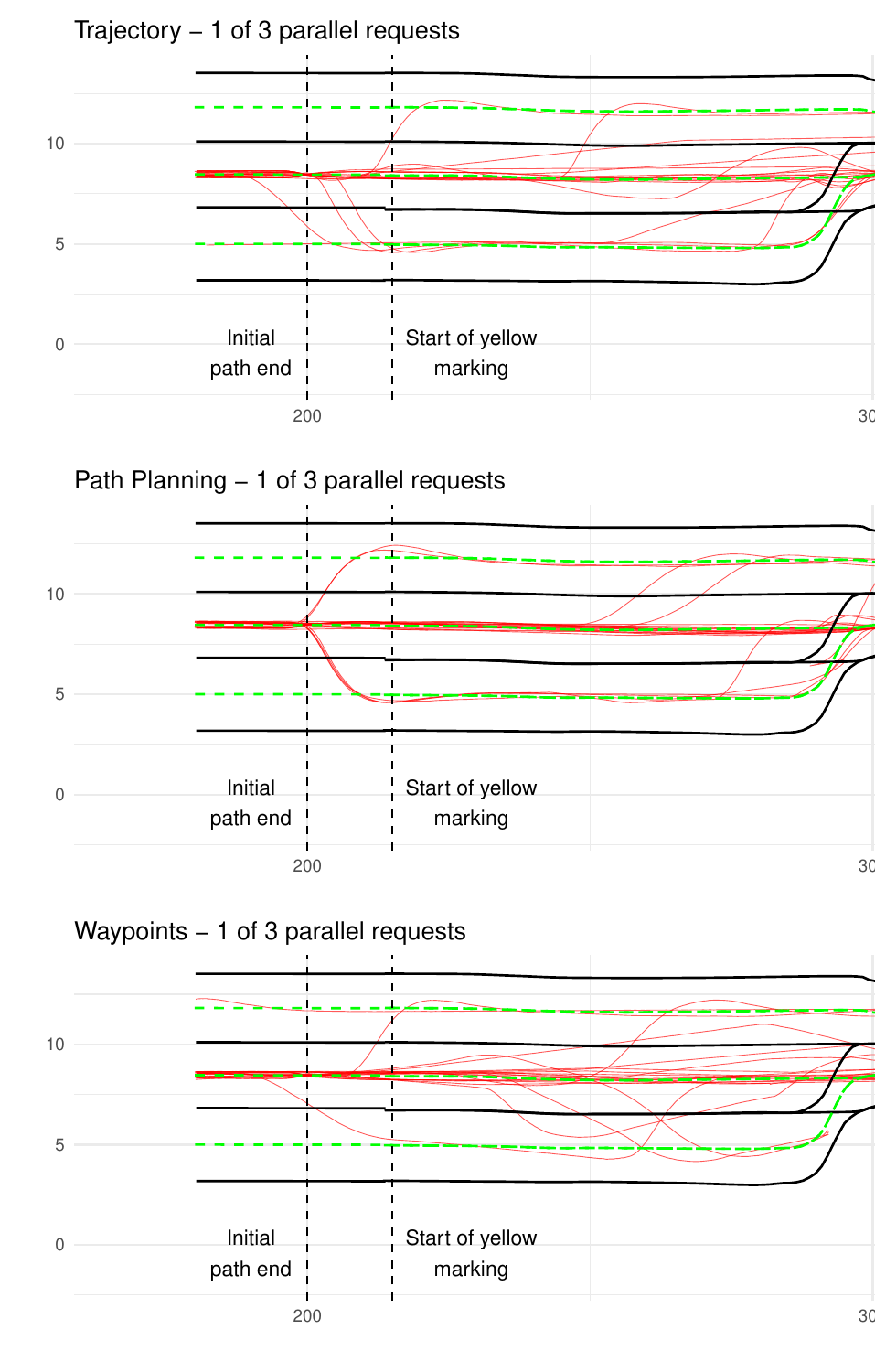}
    \caption{Three requests.}
    \label{fig:Pattern3-right}
  \end{subfigure}%
  \begin{subfigure}{0.33\textwidth}
    \centering
    \includegraphics[width=\textwidth]{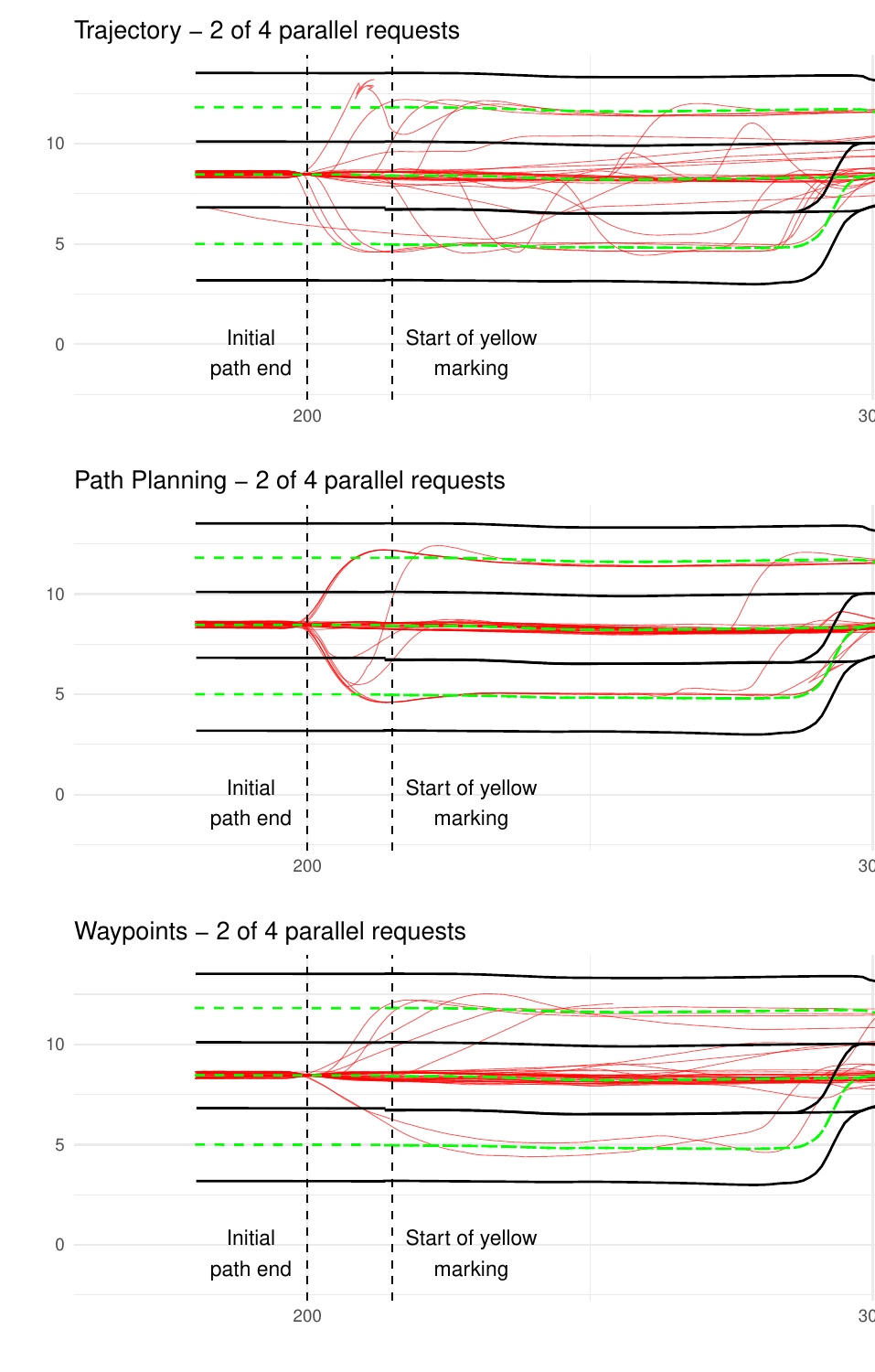}
    \caption{Four requests.}
    \label{fig:Pattern4-right}
  \end{subfigure}

    \caption{Movement patterns when approaching the right-sided road works.}
    \label{fig:startPatterns-right}
    \Description{Movement patterns when approaching the right-sided road works.}
\end{figure*}

During the study's remote assistance, the position of each requesting AV was recorded. \autoref{fig:startPatterns-left} and \autoref{fig:startPatterns-right} illustrate the AVs' paths from the end of their initial route to the start of the road works. This area was specifically selected as it contains the last overtaking position and densely represents the contrasts in the steering behavior between the \interactions. For \autoref{fig:startPatterns-right}, there are only data from two requests as for one request, the road works were always on the left side.
The complete recordings of the AV positions are visualized in the appendix \ref{app:movementPatterns}. While no statistical test was conducted, a tendency to less controlled steering can be observed for \waypoint and especially for \trajectory compared to \pathPlanning.

% left site

% right site

\subsection{Ranking and Open Feedback}

A Friedman rank sum test found a significant effect of \interaction on ranking (\chisq(2)=21.48, \pminor{0.001}, r=0.47). 
Participants clearly preferred \pathPlanning, followed by \waypoint. \trajectory was ranked the lowest. All comparisons were significant.

Users commend the system's \pathPlanning mechanism, describing it as "optimal," which suggests that the core functionality meets or even exceeds expectations. This is bolstered by the general agreement that the underlying concept of the system is solid, as one user explicitly stated, "The idea itself is good."

However, there are notable concerns regarding remote operation in general, particularly related to technical limitations such as latency, signal reception, and dead zones. These are considered potential hazards, as one user articulated, "Technical aspects such as delay, signal quality, signal dead zones, etc. could lead to more dangers than benefits." These technical issues raise questions about remote operation's general efficacy and safety, especially when vehicles navigate complex environments like construction zones. In this context, users desire more guidance, highlighting that "solutions measures for getting stuck in construction sites would have been very helpful."

Users find the \trajectory to be imprecise and time-consuming, suggesting a need for simplification and stabilization. This sentiment is captured in the statement, "Trajectories in comparison were generally too imprecise and time-consuming."

Regarding usability, for instance, users would appreciate the ability to have the vehicle move backward, as suggested, "As an improvement, I would suggest always having the option to drive backward." This is interesting as this was \textbf{always} possible.
Furthermore, there is a desire for the system to automatically focus on the most urgent cases, as mentioned in "A mode in which Main-View automatically switches to the most urgent case after a short visual warning."
Additional features that could improve the system include an automatic focus on the end of the path, as one user recommended, "When using path planning, the system could jump directly to the end of the path." Again, we actually supported this via a single click. %Users also call for better support for drawing paths, with one stating, "Since drawing is very easy, even better support would be good here"
Finally, participants mentioned the desire to be able to switch between interaction concepts.

\begin{comment}
%%%%%%% Key findings of analyzed data
% - Path planning best within each dependent variable, followed by Waypoint. Trajectory performed the worst.

% - Singular Anomalies between 4, 3, and 2 requests,  for Waypoint. Often 4 performed better than 3 (and rarely even 2) <- Average Progress, Time of first completed request, Lane Deviation, Lane Changes, Usability, Usefulness, Satisfying, and even Task Load. This is interesting.  However, we did not report all of these variables. Not sure why this is, but I suspect some sort of higher concentration due to the increased challenge.

% - One request at a time generally best regarding objective dependent variables, although no significant differences within Usability and Acceptance

% -User feedback yielded new user requirements for interactions and user interface
\end{comment}

\section{Discussion}
This work compared the three interaction techniques \trajectory, \waypoint, and \pathPlanning for remote operation of AVs outside their ODD. The interaction techniques were tested with one to four parallel requests. We found that participants preferred the \pathPlanning interaction, where they could rely most on the AVs' suggestions, which also was done in previous work~\cite{andersson2023exploring}. This leads to questions of the necessary decision support capability level, even for remote assistance.

%similarity to Jonas -> we can design for >1 requests but maybe require better systems

%Notes:
%Performance differences between manifested most strongly in the setting with four requests. Even if significant, performance differences below four were relatively minor. -> if the additional flexibility/accuracy of trajectory setting is required, it should only be applied in settings with low simultaneous workload.

%Path planning is the most restricted technique in terms of freedom of input and available vehicle actions. It is also the simplest to use as a result. Part of its performance will be induced by straight motorway scenario. The other question is whether the limited choice and interaction complexity also limited the user's SA demand (ref. SA and soundscape), enabling more focused interaction with fewer distractors or obstructing factors.

%"Ideal" number of simultaneous operations: Strong differences from four onwards would suggest 1-3 to be the ideal number of simultaneous operations. However, preference for two in terms of satisfaction suggests narrower ideal field. To discuss: one too monotonous in prolonged settings (future work), three already very demanding for sustained workload, regardless of interface / interaction technique.

%What do we do with Waypoint? It does not seem to have strong drawbacks over path planning - does it have any notable advantages (perhaps observed in the qualitative data)?

\textbf{Eye gaze}
Gaze distributions between displays are very similar across concepts. With one request, the operator focuses 100\% on the main display. With 2-4 requests, the gaze distribution is equal, with around 30\% on the secondary display and slightly above 60\% on the main display, and only a few percent on the request display. Except for one request, when the operator only needs to focus on the main display, eye gaze distributions are very similar despite an increase in the number of requests. Eye gaze distribution is also very similar across interaction concepts. This could indicate that task demand does not have an effect on how the screens are used. Also, it hints that more granular analysis may be necessary to reveal differences in eye-gaze patterns between interaction concepts.

\textbf{Number of missed requests and Lane Deviation}
Overall, one request consistently shows the best performance with the least neglected time, while Four requests consistently has the highest average neglected time and the most variability, indicating large individual differences in operator performance. The results show that the Trajectory concept stands out as a rather poor design for handling several requests simultaneously. While Path planning has a lower variation between the number of requests, the difficulty of handling requests with the Trajectory concept is accelerated as the number of requests increases. Primarily, this result indicates the importance and benefit of appropriate system capability, i.e., that lowering the task demand by automating manual tasks can help increase the number of vehicles an operator can handle simultaneously (REF). The results also show the importance of design for the task, where the Path planning concept seems to be a better design to allow for more vehicles to be controlled under a given workload.

\textbf{Neglected time analysis}
The neglected time measurements show similar patterns as the number of missed requests. As can be expected, the data indicates that an operator's ability to attend to requests in a timely manner becomes lower as the number of vehicles increases. It also shows that the interaction design affects how well (fast) the operators can respond. This has direct implications for the design of remote operation centers and the allocation of vehicles to operators. Specifically, the Trajectory concept shows a peak in neglected time for operators managing more than one vehicle, suggesting that this interaction concept may be particularly demanding, while Path planning gives an overall lower neglect time. The long error bars for operators managing four vehicles suggest high variability in how different operators handle this workload. This could be due to individual differences in capacity or working strategies stressing the need for training and good work procedures.

\subsection{How Many Parallel Requests Are Feasible For Remote Operators Or Is it even Acceptable?}
While it is sometimes implicitly assumed in the literature (e.g., by employing direct control via a steering wheel~\cite{andersson2023exploring}) that an RO will only ever directly handle one request at a time, fleet management requires supervision of several vehicles \cite{kneissl2020}, with an ever-present likelihood of several interventions being necessary simultaneously. The cognitive and motor demand on a human interacting with standard physical vehicle controls cannot necessarily be translated directly into a one-to-many (see \citet{hashimoto_human_2022}) fleet management setting and is also mediated by the mode of control. Differences in input and output modulate demand simultaneous direct intervention capabilities. Our data suggests that multiple requests are feasible within certain limits. We even found that satisfaction was higher with two than with only one request (see Section~\ref{sec:acceptance}). Additionally, no clear patterns are observable that indicate that a higher number of parallel requests leads to higher variability in the movement patterns (see \autoref{fig:startPatterns-left} and \autoref{fig:startPatterns-right}). The boundaries arise more from the question of how long AVs can be neglected and how many missed requests are acceptable or can be handled by other ROs? This emphasizes the need to also consider organizational aspects in the design of remote operation systems. \autoref{fig:missingRequests} and \autoref{fig:avgMax_interaction} show initial data on what to expect with different techniques and number of parallel requests with \textbf{alert and refreshed} ROs. The question remains whether this can be generalized to all contexts or whether there are configurations (driving environment, vehicle types, traffic density) that put increased demand on input or output capabilities and result in lower scalability potential.

\subsection{System Capability Necessities for Remote Operation}
% does this mean that path planning is one of the most relevant capabilities needed until full AVs are available?
This study found that participants could handle more than one request in parallel. While no sustained workload was induced over longer periods of time, at least for two requests, the relevant dependent variables were still in acceptable ranges. For the "ideal" number of simultaneous operations, data shows a noticeable divergence when more than three operations are involved. This suggests that 1-3 operations may be an appropriate range. However, user satisfaction peaks at two operations, narrowing the ideal range further. Future work should explore whether one operation becomes monotonous over time and whether managing three operations becomes overly demanding, irrespective of the interface or interaction technique used.

However, we highlight that the appropriateness of multiple request-handling depends not only on the interface for the RO but heavily on the capabilities of the AV controlled. 
Path planning is the most constrained technique in terms of input flexibility and available vehicle actions, making it the simplest to deploy for an operator. However, it requires the highest technical sophistication on the AV side. The technical capabilities of the AV necessary do not alter between \trajectory and \waypoint as a \trajectory is simply multiple waypoints taken together. The question then becomes whether it offers any distinct advantages over the reduced technical capabilities of AVs, possibly reflected in qualitative data. There is a potential for improving overall effectiveness and usability if both modes are available.

%- Simulator setup - pros and cons (notes)
%+ mostly simple to understand simulator keeps the focus on compared modalities and reduced additional complexity
%+ single monitor setup avoided further distraction (maybe on the cost of immersion)
%+ One study run always was conducted in less than 90 minutes -> participant attention span was not too strained
%+ Supportive features included in the HMI that went further than the most basic implementation made the comparison more realistic and valid

%- Higher immersion may lead to more realistic usage (especially concerning trust in ADS and path suggestions)
%- More additional HMI features, not directly related to the modalities, would increase immersion -> e.g. demo overview map of vehicles, while no request is active
%- used scenario only reflects comparison on linear and rather perfect road problem situation. -> also implies steep learning effects, since participants already know the scenario and how to solve it at some point
%- Supportive features for modalities and HMI were freely selected without prior knowledge of which are best and which may be missing
%- no quick generalized "delete current path and stop vehicle" function

% external validity - different scenarios?
% dependent variables - subjective and objective lane deviation , neglected time! How much neglect is bearable / acceptable (context dependency) --> KPI
%

% focused on remote assistance - no monitoring

\subsection{The Ideal Interaction Mode and the Need for Alternatives}
Remote assistance is a multi-faceted interaction challenge. Therefore, the simulator to study parts of it must reflect these. Our simulator for remote vehicle operation provided simplicity and focus, allowing participants to compare operational modes easily. For this purpose, the three modalities were presented as exclusive alternatives, assuming ideal operating conditions. \pathPlanning received the \textbf{best} rating in \textbf{all} dependent variables, thereby, showing a clear preference. However, for \pathPlanning specifically, our implementation meant that the operation was predicated on there being at least one optimal or correct path to choose from. In a realistic setting, the possibility of system deficiencies leading to suboptimal path recommendations would need to additionally be considered, putting \pathPlanning as the lone preferred mode into perspective. Thus, \pathPlanning can be seen as a reasonable default but with a need for a workaround in case the default is not viable. On the basis of the results from this study, the workaround option would be the \waypoint due to possessing the necessary capabilities to more finely define a route while performing better overall than \trajectory. The eventual interaction concept that enables both scalability and finetuning- or workaround-options is one with \pathPlanning as the default control mode and waypoint as the supplementary mode.%Due to this focus, each study was run under 90 minutes, maintaining participant attention. The HMI also included extra supportive features, adding realism and validity to the study.

An additional consideration in particular towards \pathPlanning is its active recommender-characteristic, which renders it most convenient. This raises additional questions regarding accountability or liability in case its recommendations are wrong~\cite{NARAYANAN2020255}. Who would be held responsible if a fleet management system were to cause accidents based on suboptimal trajectory recommendations? Is it a system defect or a failure of the human operator, who should perhaps have verified the trajectory? Is it the responsibility of the third party offering the fleet management software or the OEM's who offered it to the fleet owner as part of a teleoperation-ready vehicle package? %multiple requests in parallel are possible. However, novel questions arise, for example: what happens when the AV suggests wrong trajectories? Who is responsible? Can remote operation, therefore, not be provided by third parties but must be handled by the vendor of the AV? Is this a legal question?
While we cannot determine these legal requirements, our study suggests that if allowed, \pathPlanning is an appropriate interaction mechanism for ROs.

\subsection{Methodology}
Studying remote operation is difficult due to the numerous activities, scenarios, and potentially confounding factors. Therefore, setups need to be specifically fashioned towards the specific investigative focus (vehicle control, status monitoring, routing, etc.), with one-size-fits-all setups not being feasible nor appropriate in most instances. In our case, this was the case with the focus on the interaction possibilities \pathPlanning, \waypoint, and \trajectory, which we decided to investigate in a comparative, single-view setting. Adding, for example, additional monitors with other information, such as a map, would have been easily doable and would be a sensible extension in a follow-up to explore a combination with spatial planning or traffic load management. However, this would reduce the focus on the intervention.
The multifacetedness of remote interaction is also reflected in relevant dependent variables. We logged numerous data and also incorporated usability, task load, and acceptance as subjective measures. Arguably, the ``neglected time'' was the most important efficiency-related measure for the focus on parallel requests. This metric also has high relevance for real implementations and could serve as a KPI (key performance indicator) for overall performance.

\subsection{Limitations}
We studied remote operations with N=23 rather young participants. While this limits generalizability, our data can be seen as a baseline of how remote operation could work in a best-case scenario. Nonetheless, data on prolonged exposure and studies with older participants are necessary. Additionally, due to the focus on high internal validity, we did not have a full setup that could be envisioned for ROs. Our results should, therefore, be replicated in such a setup. Finally, we focused on one scenario. Other scenarios must be evaluated in the future. 

The single-monitor setup minimizes distractions but may limit immersion and may not represent an actual operator environment with multiple monitors (e.g., see \citet{schrank2024human}). Nonetheless, such a setup was, for example, also used by \citet{andersson2023exploring}. \citet{kettwich_teleoperation_2021}, who supplied 7 monitors to participants, even reports that participants stated that one monitor would suffice. The major limitation for external validity, which was, however, needed for internal validity, is that the scenario focuses on ideal road conditions with a rather linear street, narrowing the study's applicability and risking steep learning curves. We reduced this by also mirroring the road works. However, this is still a limitation. Further, there is a limitation to the fidelity of both the visual and physical simulation in the apparatus. We used the Polarith AI asset\footnote{\url{https://assetstore.unity.com/packages/tools/behavior-ai/polarith-ai-pro-movement-with-3d-sensors-71465}; Accessed on 08.12.2024} to simulate the AV and the Mobile Traffic System asset\footnote{\url{https://assetstore.unity.com/packages/tools/behavior-ai/mobile-traffic-system-194888}; Accessed on 14.08.2024} to simulate traffic. While this does not represent the real world, no participant mentioned any confusion about the simulation. Lastly, the lack of a quick "delete and stop" function is a notable omission regarding already started trajectories. Further features would be an overview map of vehicles while no request is active.

In this work, relevant factors such as latency~\cite{zhao2024remote} or potential additional feedback for the RO~\cite{zhao2024remote} were out of scope. Future work should investigate its effects and potential countermeasures (e.g., for latency, there are possibilities to reduce bandwidth or robust prediction algorithms. See \citet{zhao2024remote} for an overview).

Furthermore, the driving experience could have affected performance in our task. Future work could consider this in further evaluations.

Finally, some of our assumptions might not hold for realistic implementation (e.g., the maximum 90° internal angle for waypoints might need to be reduced). Implementations using real vehicles will need to evaluate this in the future.

\section{Conclusion}
This work compared three interaction techniques for the remote operation of AVs: \pathPlanning, \waypoint, and \trajectory. We tested these with one to four parallel requests. The designed simulator can be re-used for multiple future use cases and will be provided to interested researchers. In a within-subjects study with N=23 participants, we found that \pathPlanning was clearly preferred and led to the best overall results. Interestingly, participants were most satisfied with two parallel requests, showing that multiple parallel requests, at least under study conditions, are possible. We discuss these findings in light of necessary AV capabilities and considerations for multiple request handling. Our work helps to safely introduce AVs into general traffic despite potential technical limitations.

\section*{Open Science}
The Unity scenario will be available upon request. This includes installation instructions and information on required 3rd party Unity assets. Anonymized data and evaluation scripts are available at \url{https://github.com/M-Colley/roads-chi25-data}. 
%For the review, we anonymously provide the application via: \url{https://anonymous.4open.science/r/BuildsForROADS-61C1/README.md}.

\begin{acks}
We thank all study participants. Mark Colley would also like to thank Johannes Schöning for the invitation to the supportive \textit{CHITogether} writing retreat.
\end{acks}

\bibliographystyle{ACM-Reference-Format}
\bibliography{sample-base}

\appendix

\section{Appendix - Movement Patterns}
\label{app:movementPatterns}

% Splitting idea taken from: https://tex.stackexchange.com/questions/278727/split-subfigures-over-multiple-pages
\begin{figure*}[ht!]
\centering
\small
    \begin{subfigure}[c]{\linewidth}
        \includegraphics[width=\linewidth]{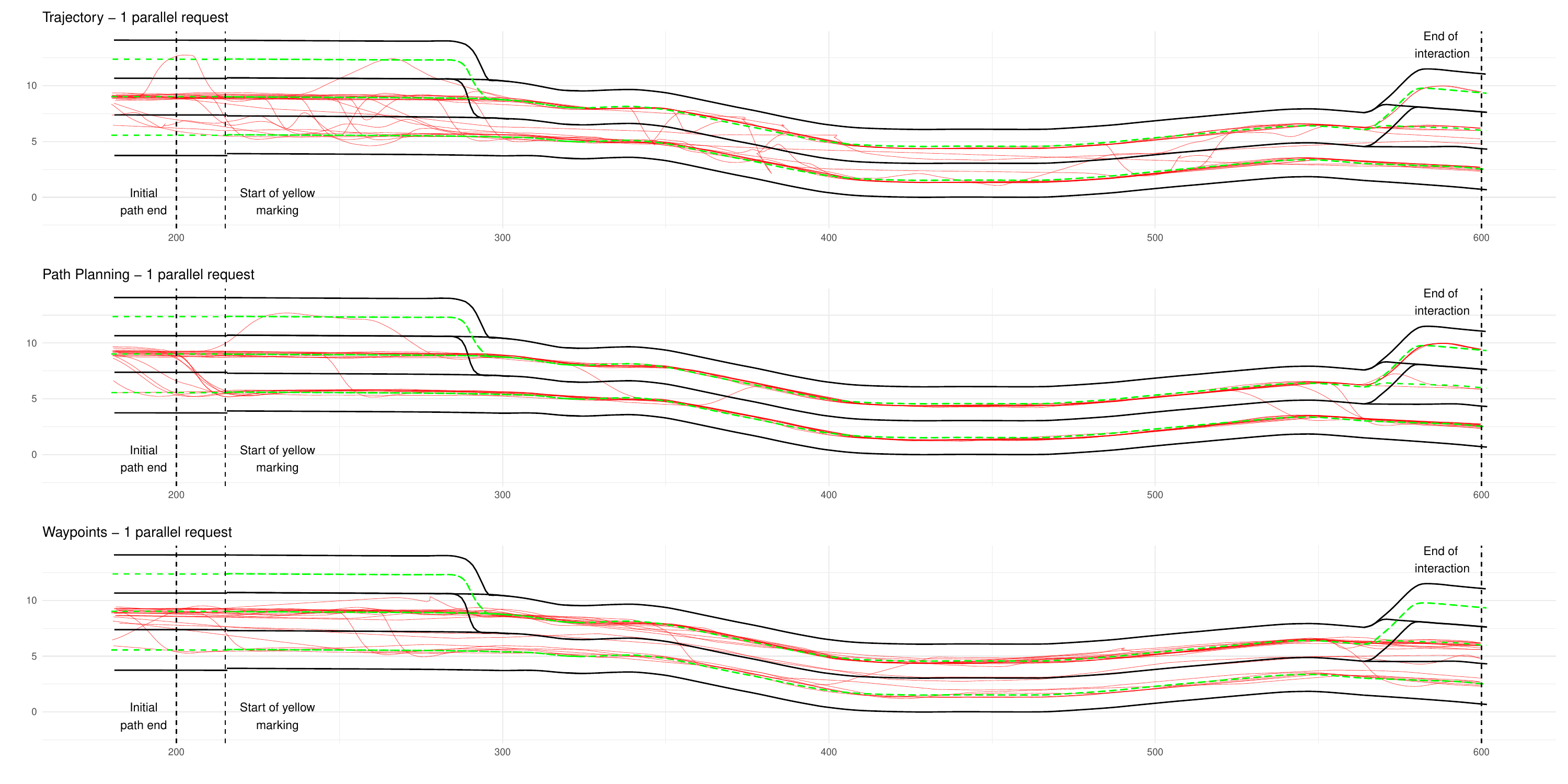}
        \caption{One request.}\label{fig:movement-1}
        \Description{}
    \end{subfigure}
    
    \begin{subfigure}[c]{\linewidth}
        \includegraphics[width=\linewidth]{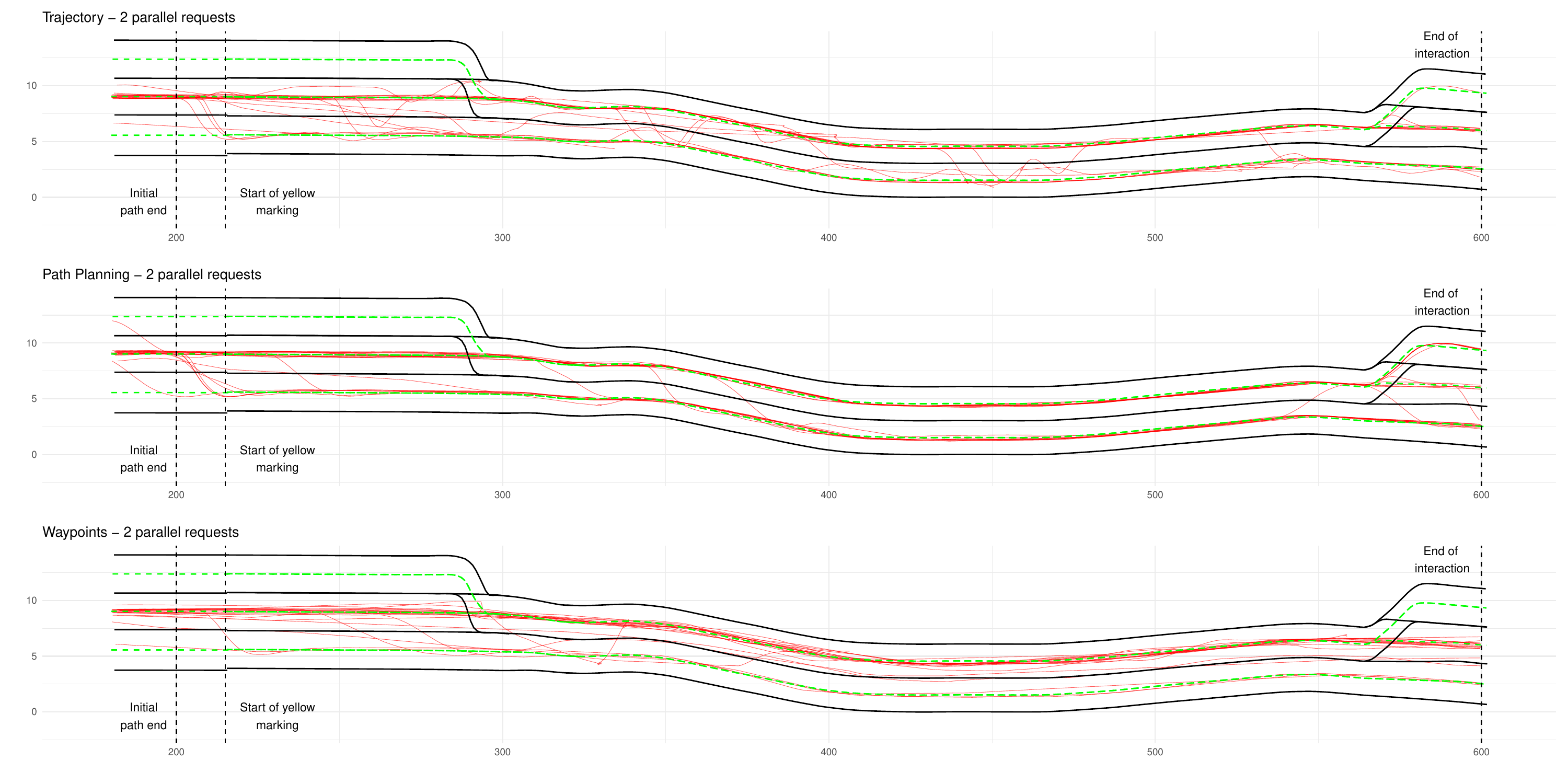}
        \caption{Two parallel requests.}\label{fig:movement-2}
        \Description{}
    \end{subfigure}
    %\caption{.}
   \Description{}
\end{figure*}
\begin{figure*}[ht!]\ContinuedFloat
        \begin{subfigure}[c]{\linewidth}
        \includegraphics[width=\linewidth]{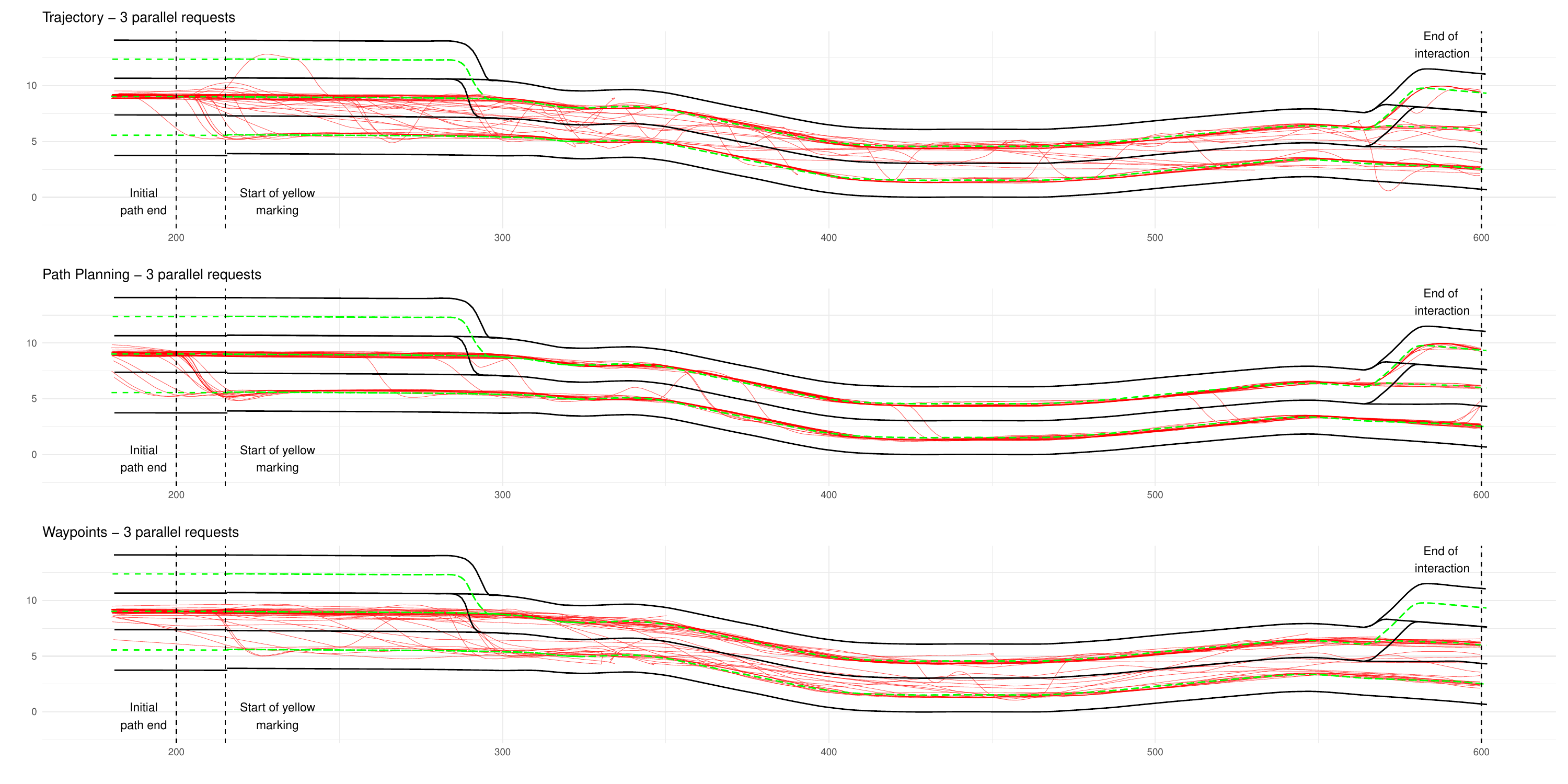}
        \caption{Three parallel requests.}\label{fig:movement-3}
        \Description{}
    \end{subfigure}
    
        \begin{subfigure}[c]{\linewidth}
        \includegraphics[width=\linewidth]{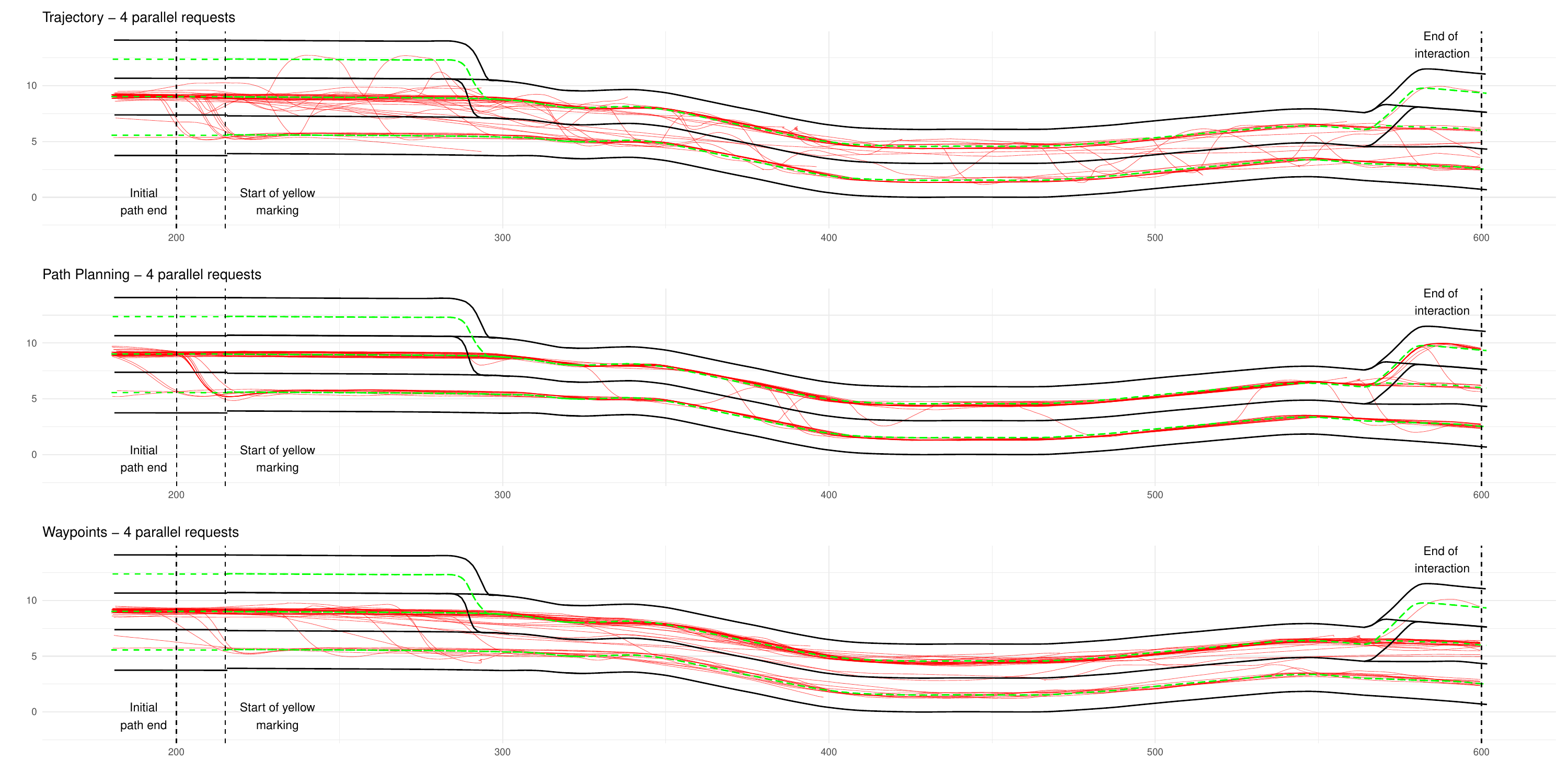}
        \caption{Four parallel requests.}\label{fig:movement-4}
        \Description{}
    \end{subfigure}
    \caption{Overview of movement patterns for the left-sided road works.}
    \label{fig:overview-movement}
   \Description{Overview of movement patterns for the left-sided road works.}
\end{figure*}

\begin{figure*}[ht!]
\centering
\small
    \begin{subfigure}[ht!]{\linewidth}
        \includegraphics[width=\linewidth]{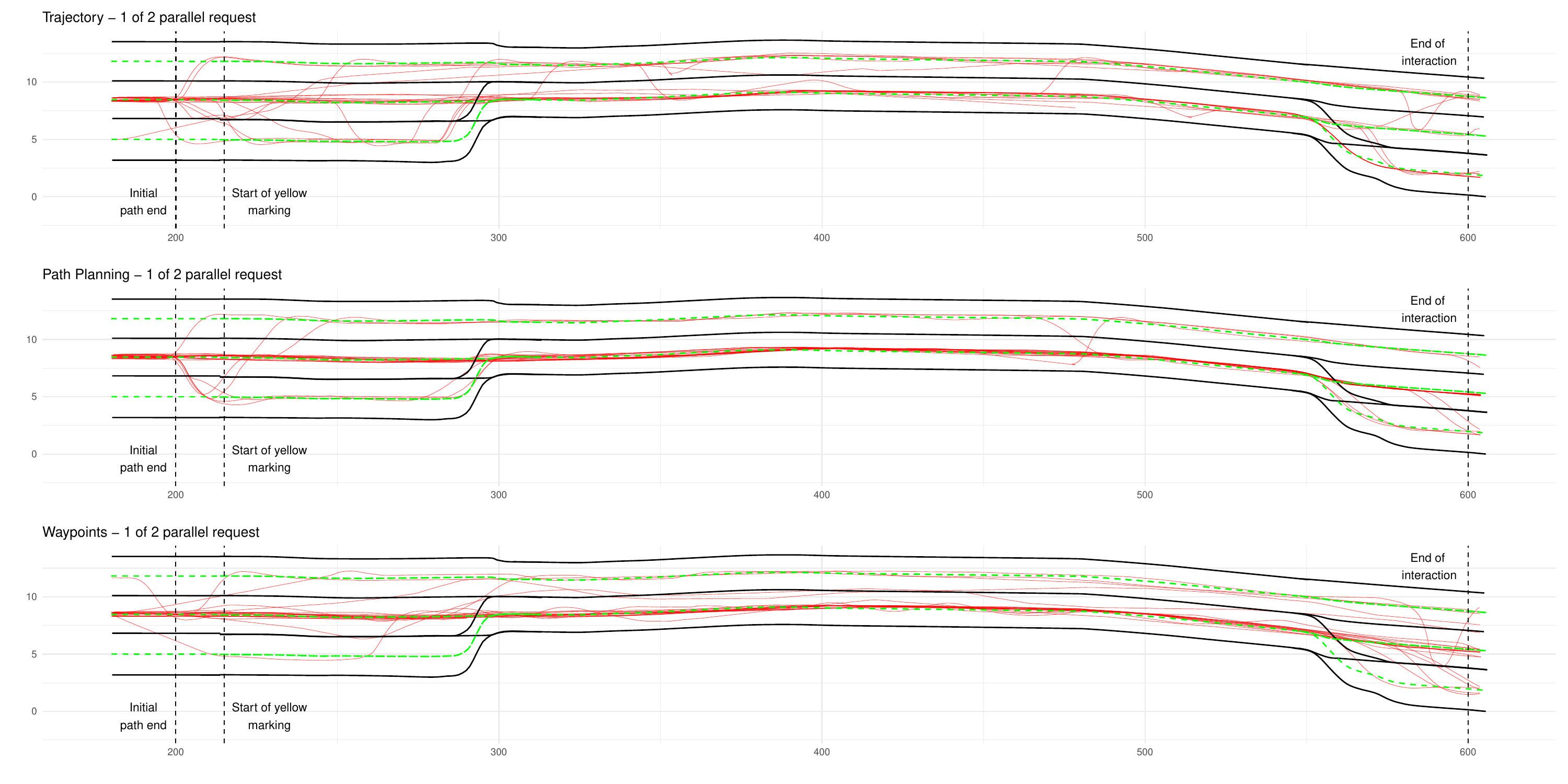}
        \caption{Two parallel requests.}\label{fig:Rmovement-1}
    \end{subfigure}
    \begin{subfigure}[ht!]{\linewidth}
        \includegraphics[width=\linewidth]{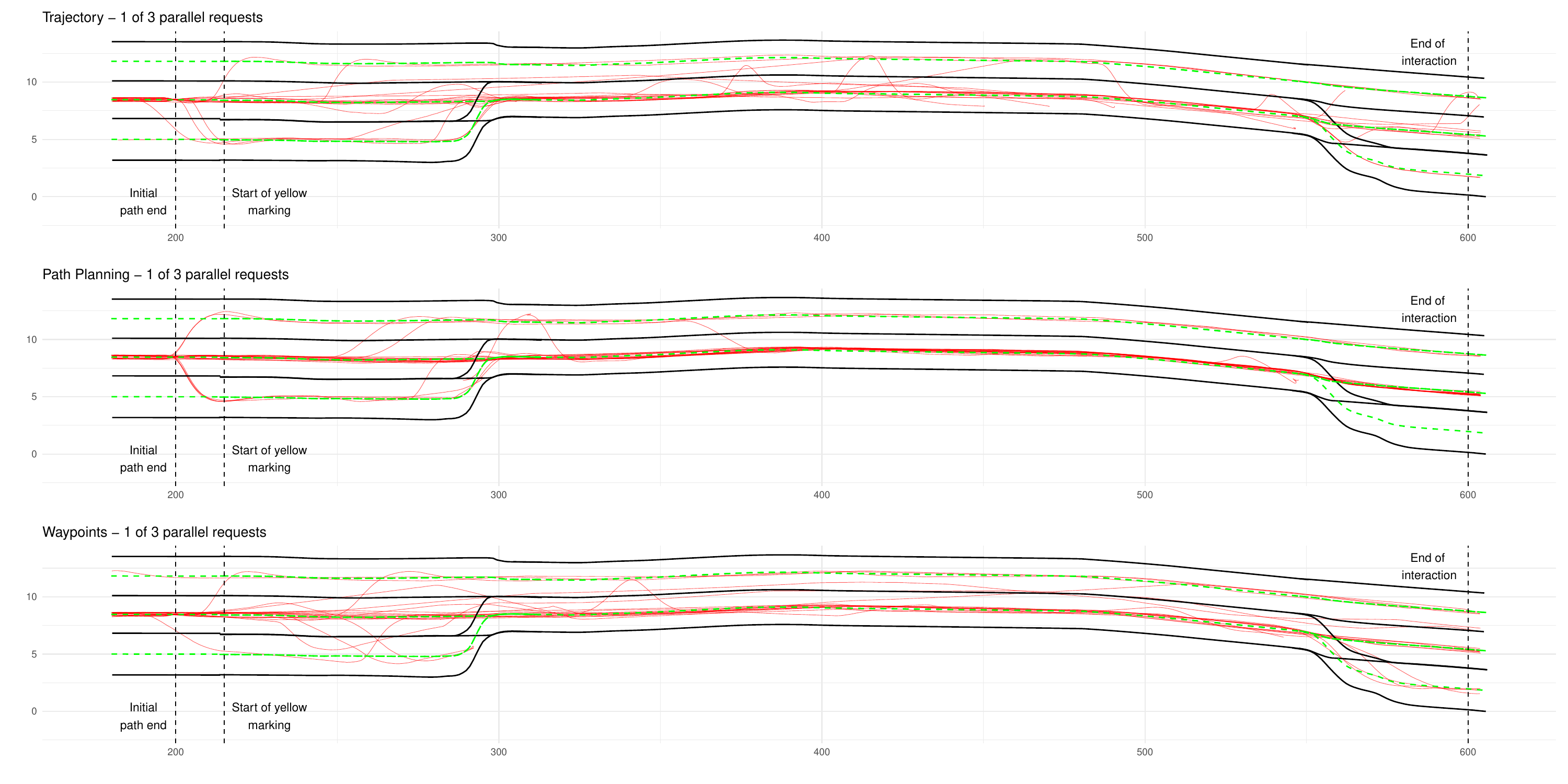}
        \caption{Three parallel requests.}\label{fig:Rmovement-2}
    \end{subfigure}
     \Description{Overview of movement patterns for the right-sided road works.}
\end{figure*}

\begin{figure*}[ht!]\ContinuedFloat
    \begin{subfigure}[ht!]{\linewidth}
        \includegraphics[width=\linewidth]{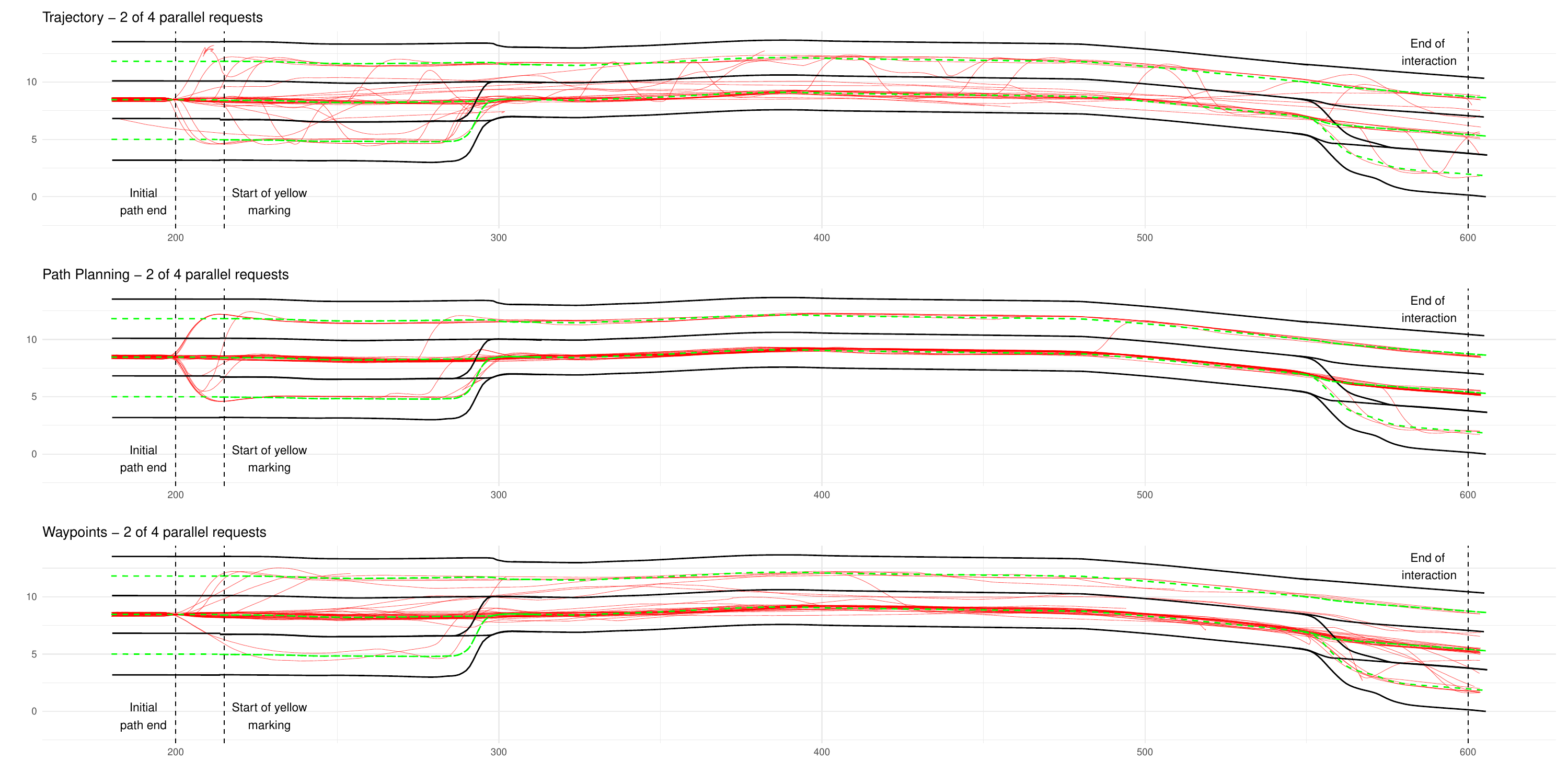}
        \caption{Four parallel requests.}\label{fig:Rmovement-3}
    \end{subfigure}
    \caption{Overview of movement patterns for the right-sided (mirrored) road works.}
    \label{fig:overview-Rightmovement}
    \Description{Overview of movement patterns for the right-sided road works.}
\end{figure*}

\end{document}